\documentstyle[12pt]{article}   

\input psbox

\newcommand{\be}{\begin{equation}}
\newcommand{\ee}{\end{equation}}
\newcommand{\ba}{\begin{eqnarray}}
\newcommand{\ea}{\end{eqnarray}}
\newcommand{\bfig}{\begin{figure}}
\newcommand{\efig}{\end{figure}}
\newcommand{\prt}[1]{\left( #1 \right)}
\newcommand{\cld}[1]{\left[ #1 \right]}

\newcommand{\bfrac}[2]{{\displaystyle\frac{#1}{#2} }}
\newcommand{\sfrac}[2]{{\scriptstyle\frac{#1}{#2} }}
\newtheorem{theo}{Theorem}
\newtheorem{conj}{Conjecture}

\newcommand{\MW}{m_W}
\newcommand{\MZ}{m_{Z}}
\newcommand{\Mho}{m_{h^0}}
\newcommand{\MHo}{m_{H^0}}
\newcommand{\MAo}{m_{A^0}}
\newcommand{\MH}{m_{H^{\pm}}}
\newcommand{\ab}{\beta-\alpha}
\newcommand{\sw}{ \sin\theta_W}
\newcommand{\cw}{ \cos\theta_W}

\newcommand{\sws}{ \sin^2\theta_W}
\newcommand{\cws}{ \cos^2\theta_W}

\newcommand{\enertot}[1]{ {\cal E}_{#1} }

\newcommand{\vcpsboxto}[3]{{$\vcenter{{\psboxto(#1;#2){#3}}}$}}

\newcommand{\rfa}{ \mbox{\vcpsboxto{3.5cm}{0cm}{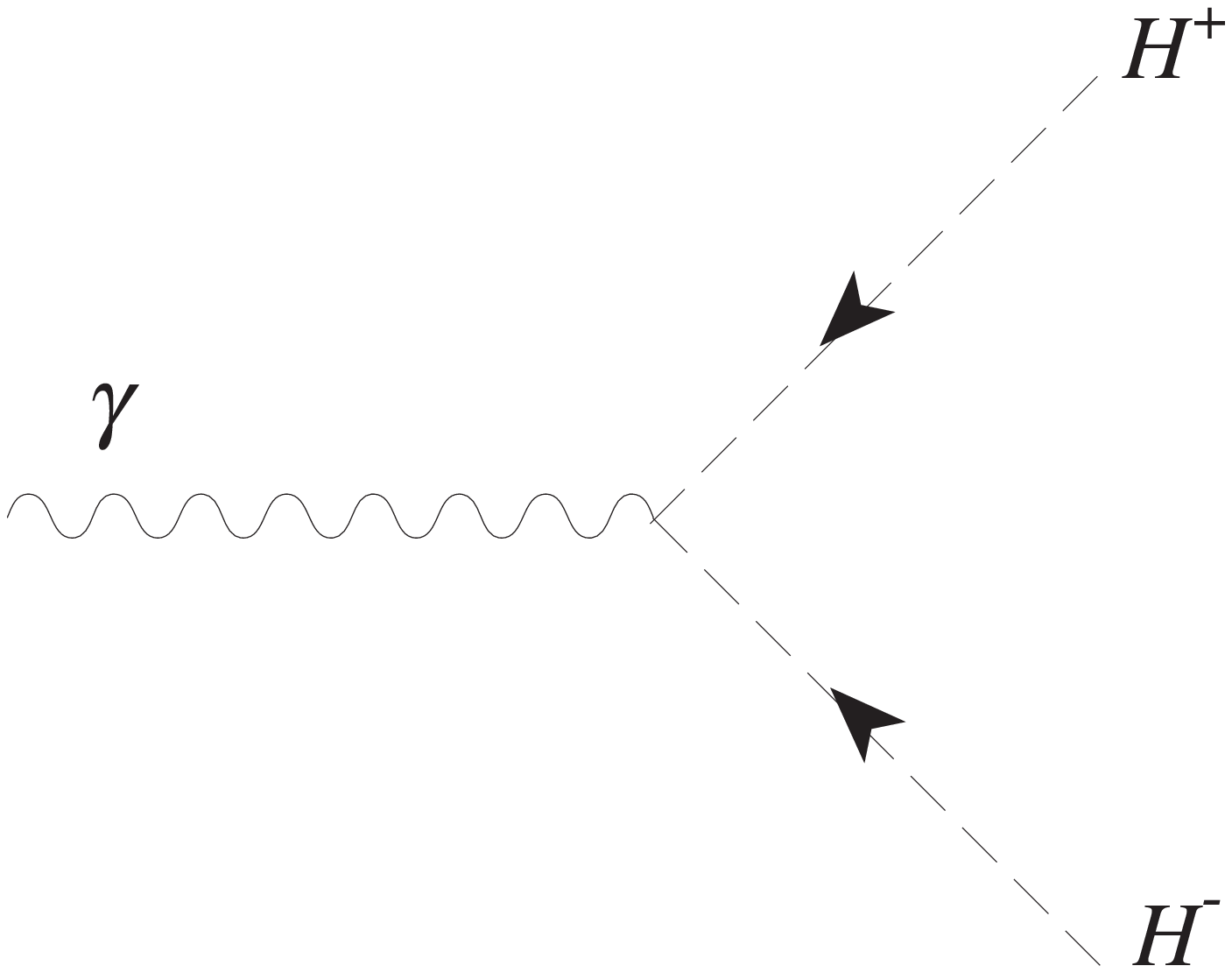}}}
\newcommand{\rfb}{ \mbox{\vcpsboxto{3.5cm}{0cm}{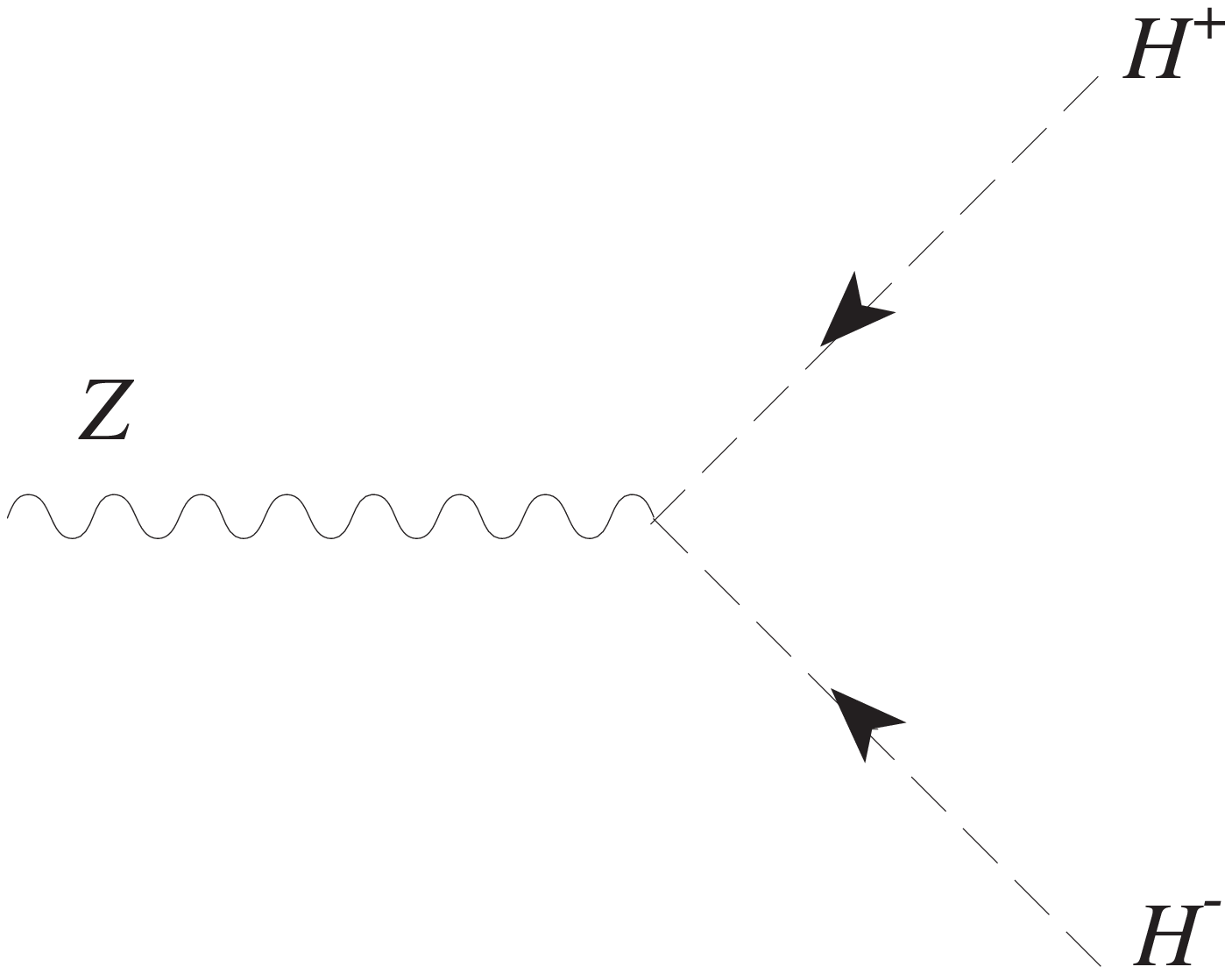}}}
\newcommand{\rfc}{ \mbox{\vcpsboxto{3.5cm}{0cm}{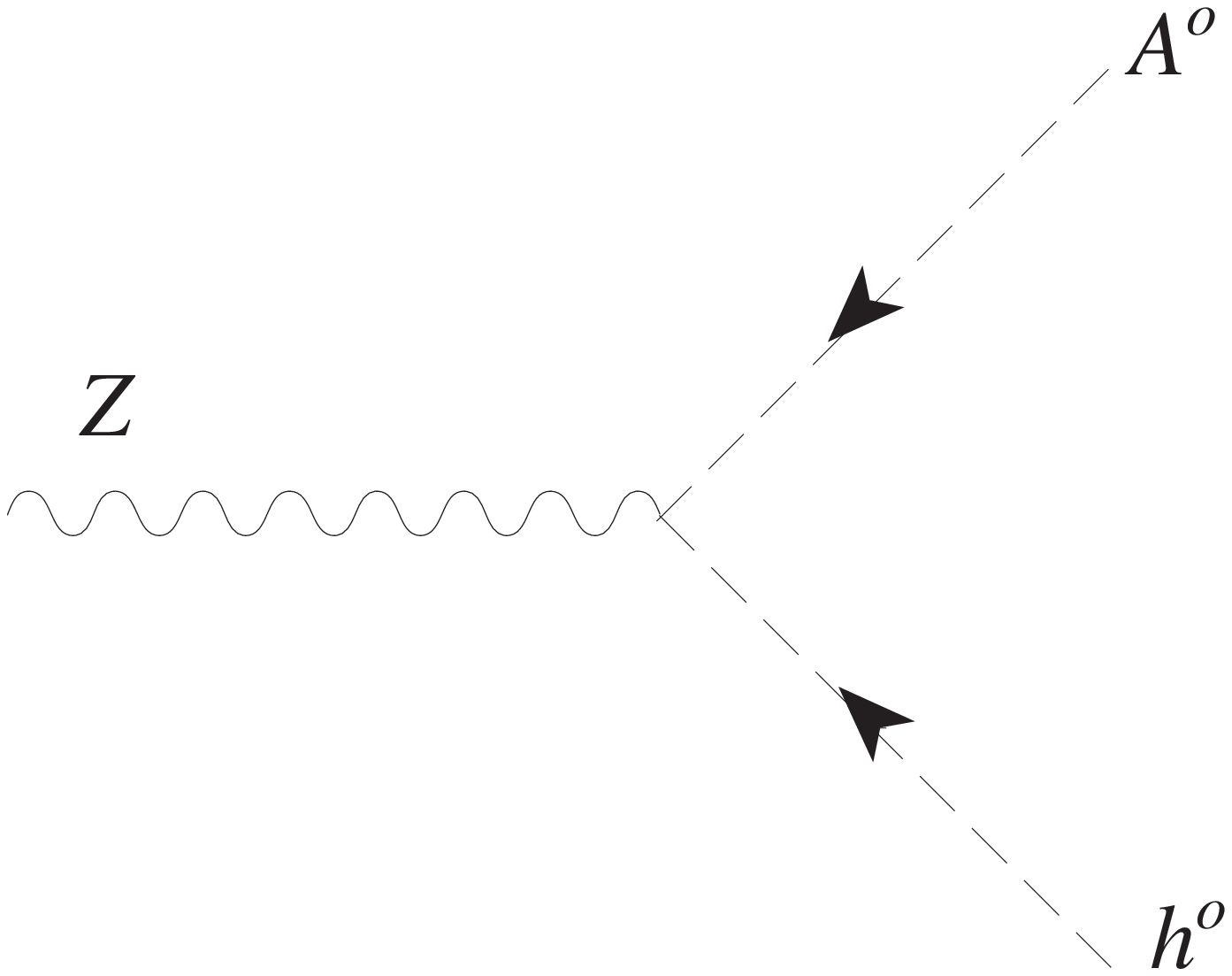}}}
\newcommand{\rfd}{ \mbox{\vcpsboxto{3.5cm}{0cm}{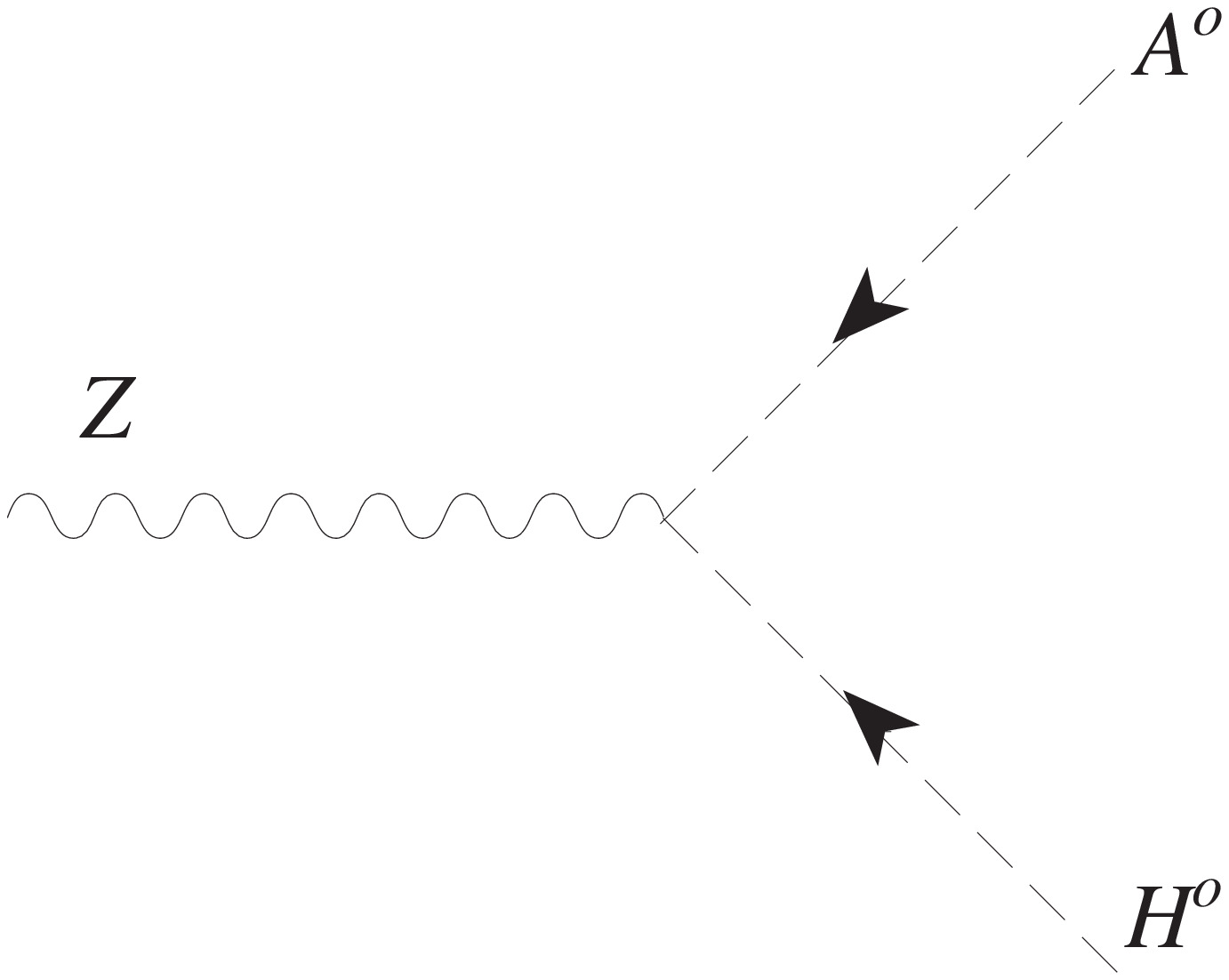}}}
\newcommand{\rfe}{ \mbox{\vcpsboxto{3.5cm}{0cm}{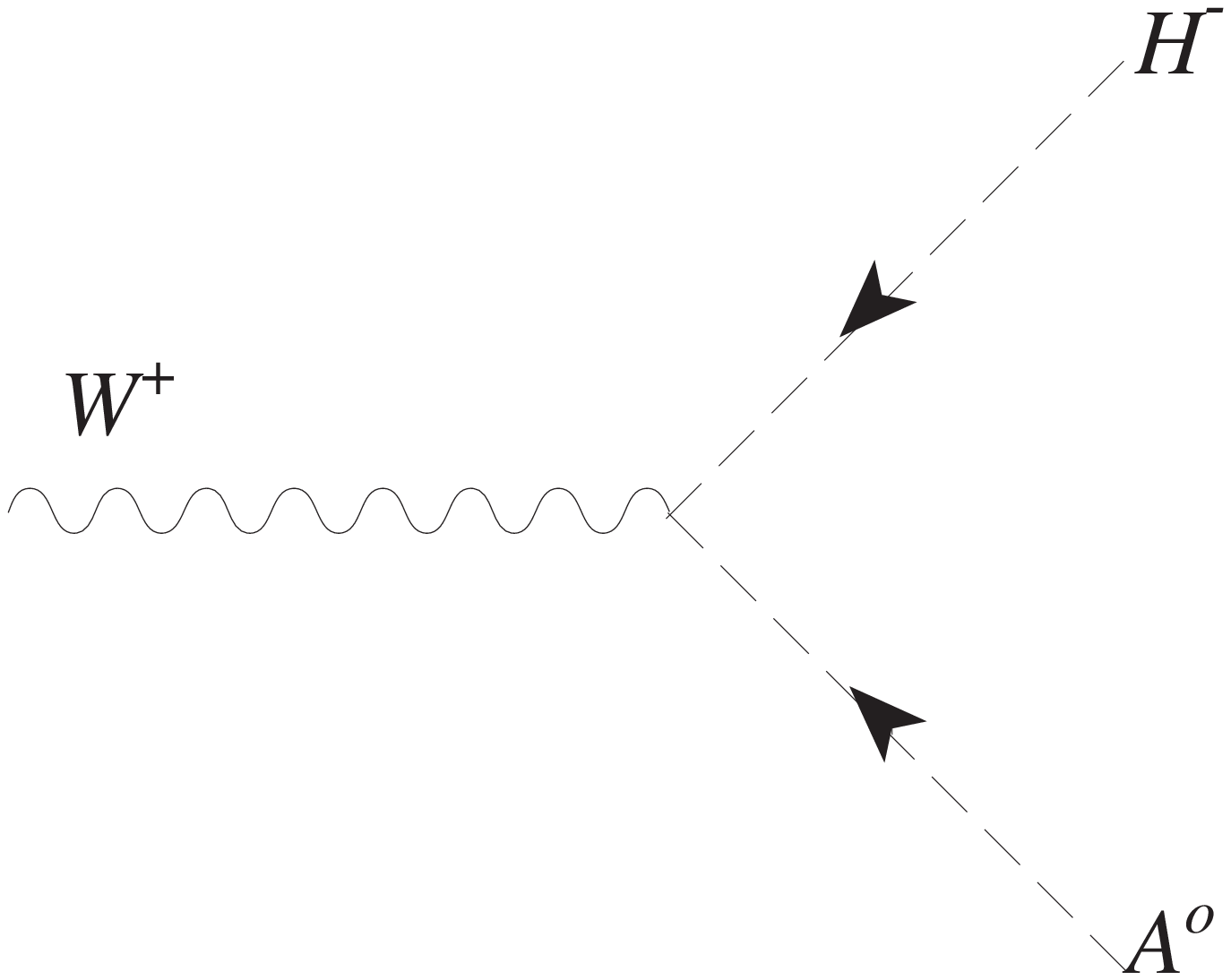}}}
\newcommand{\rff}{ \mbox{\vcpsboxto{3.5cm}{0cm}{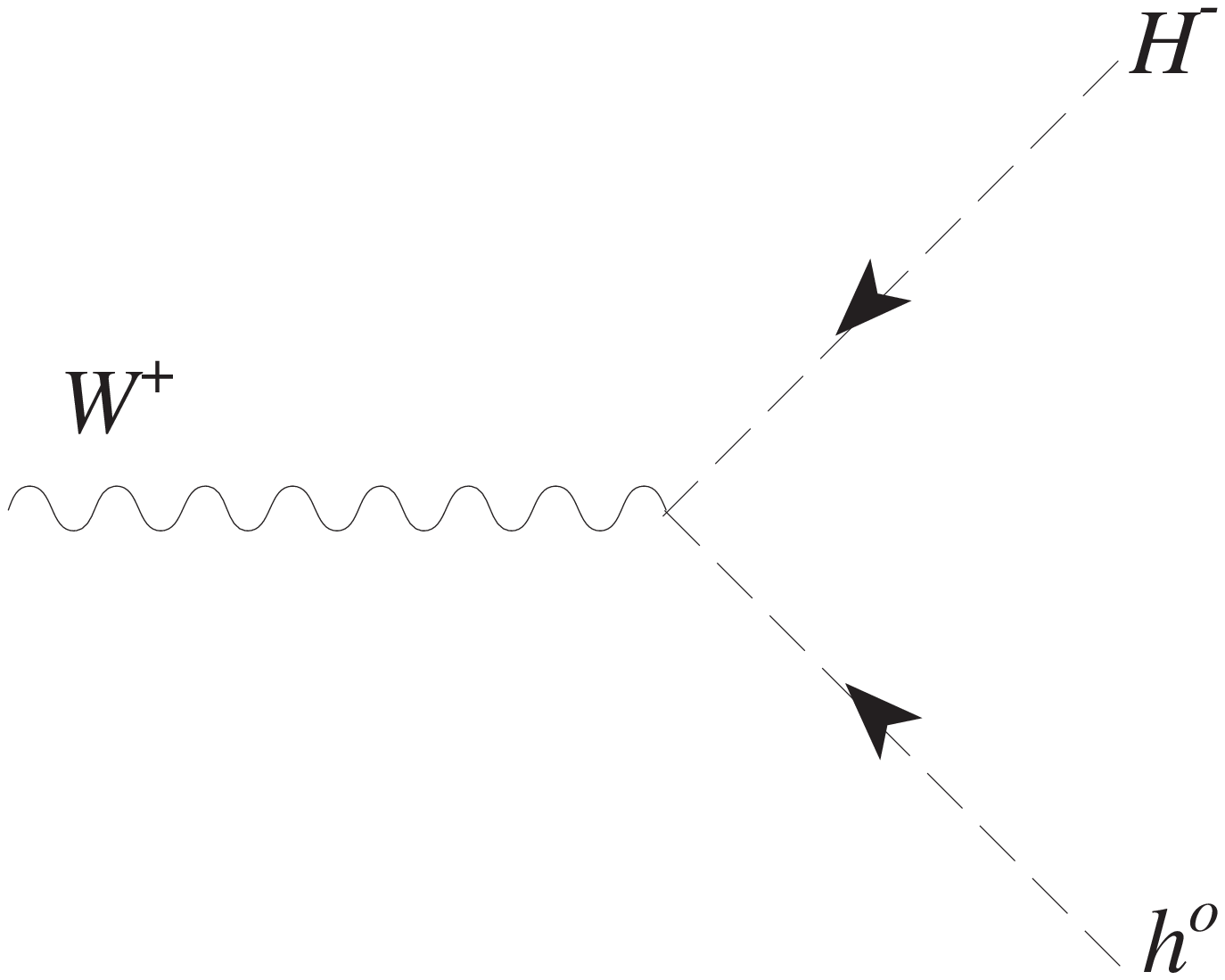}}}
\newcommand{\rfg}{ \mbox{\vcpsboxto{3.5cm}{0cm}{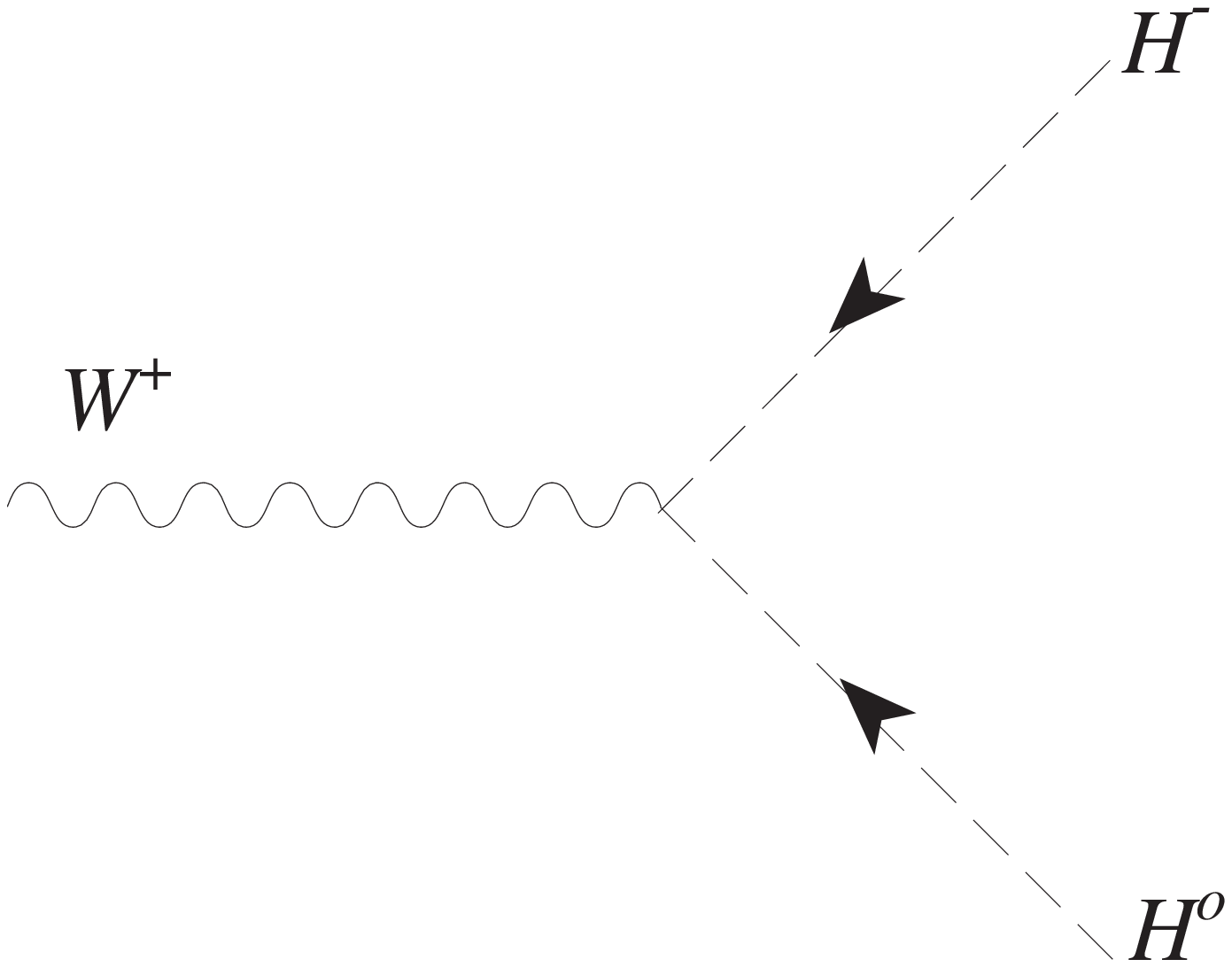}}}
\newcommand{\rfh}{ \mbox{\vcpsboxto{3.5cm}{0cm}{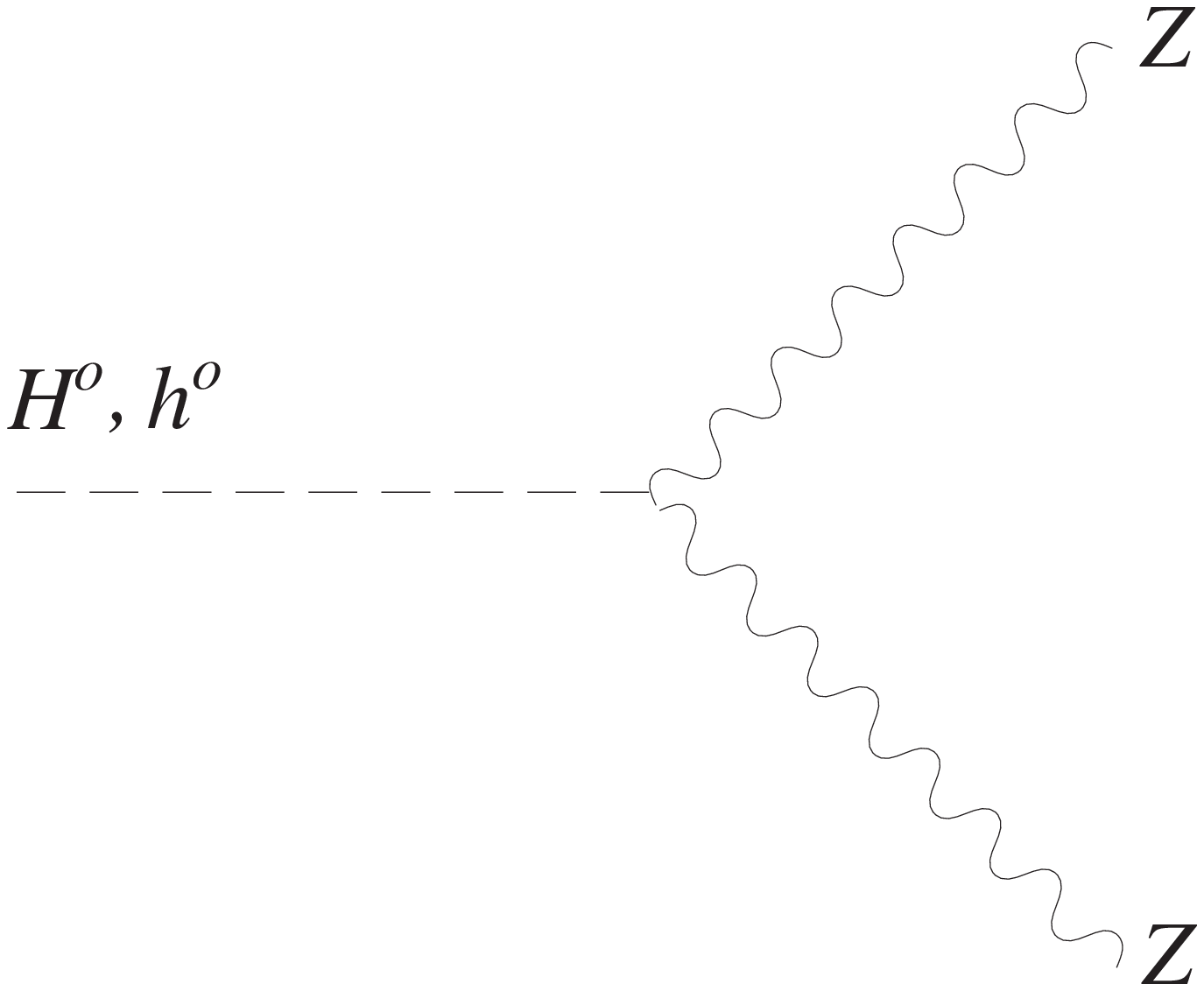}}}
\newcommand{\rfi}{ \mbox{\vcpsboxto{3.5cm}{0cm}{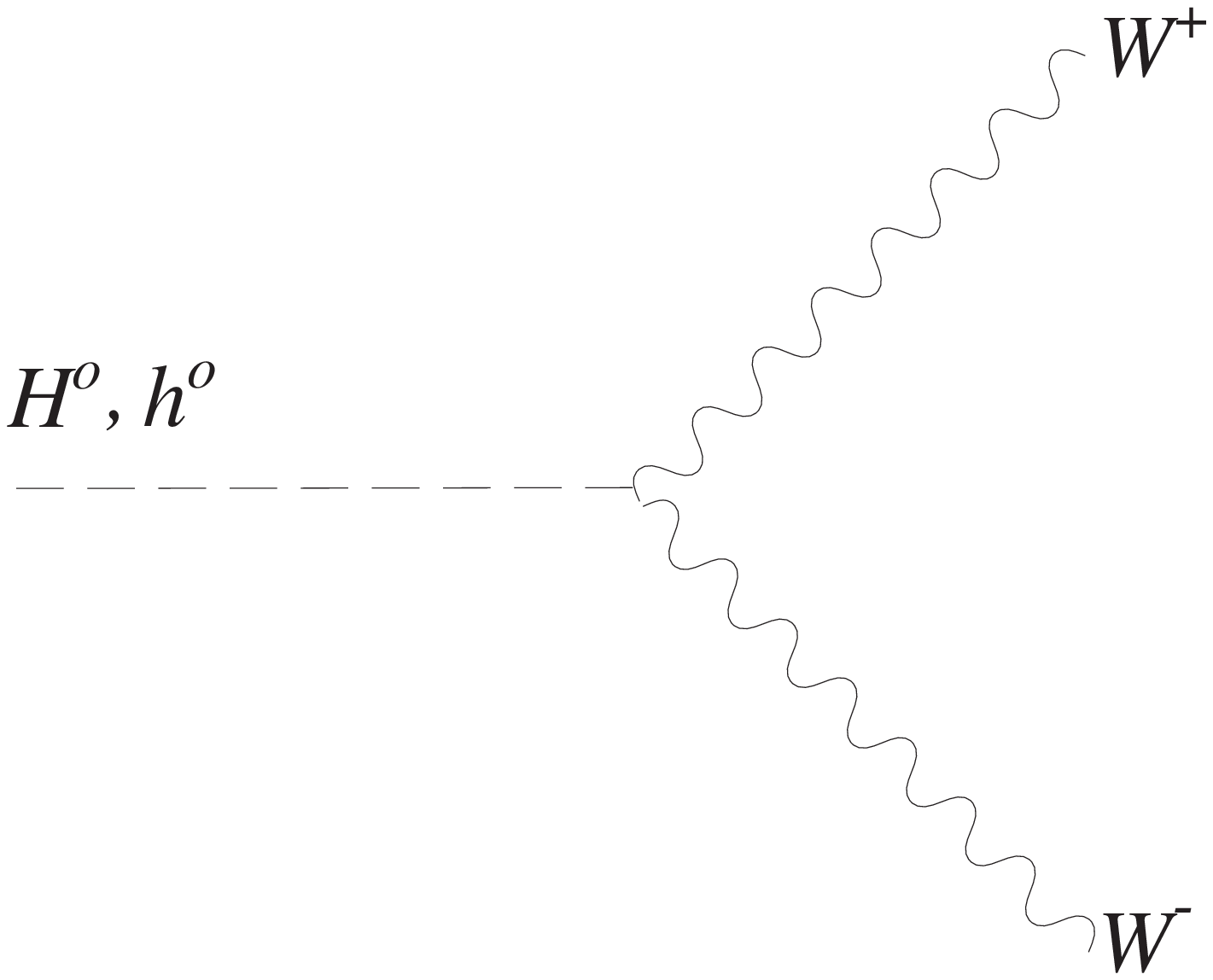}}}
\newcommand{\rfj}{ \mbox{\vcpsboxto{3.5cm}{0cm}{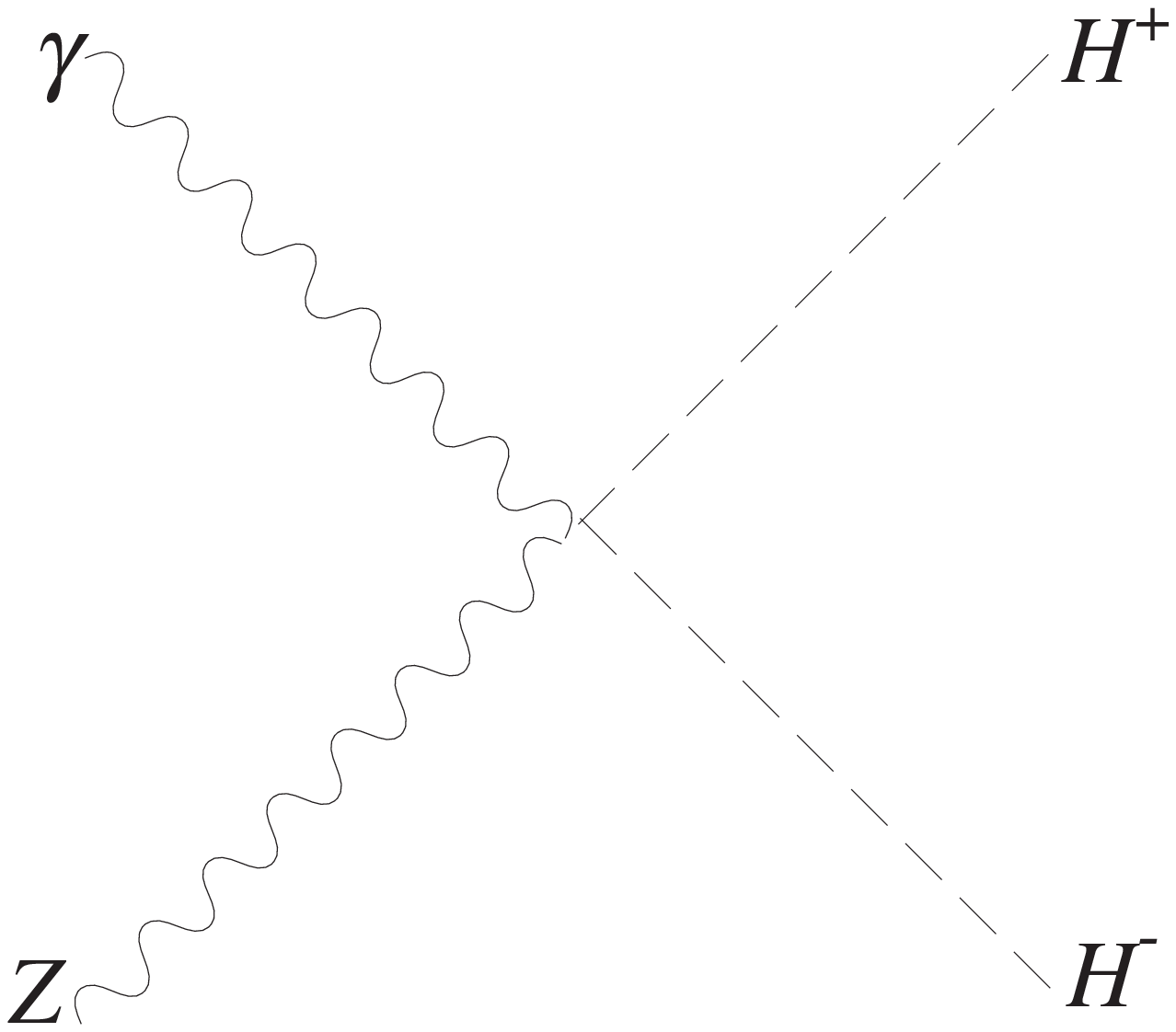}}}
\newcommand{\rfk}{ \mbox{\vcpsboxto{3.5cm}{0cm}{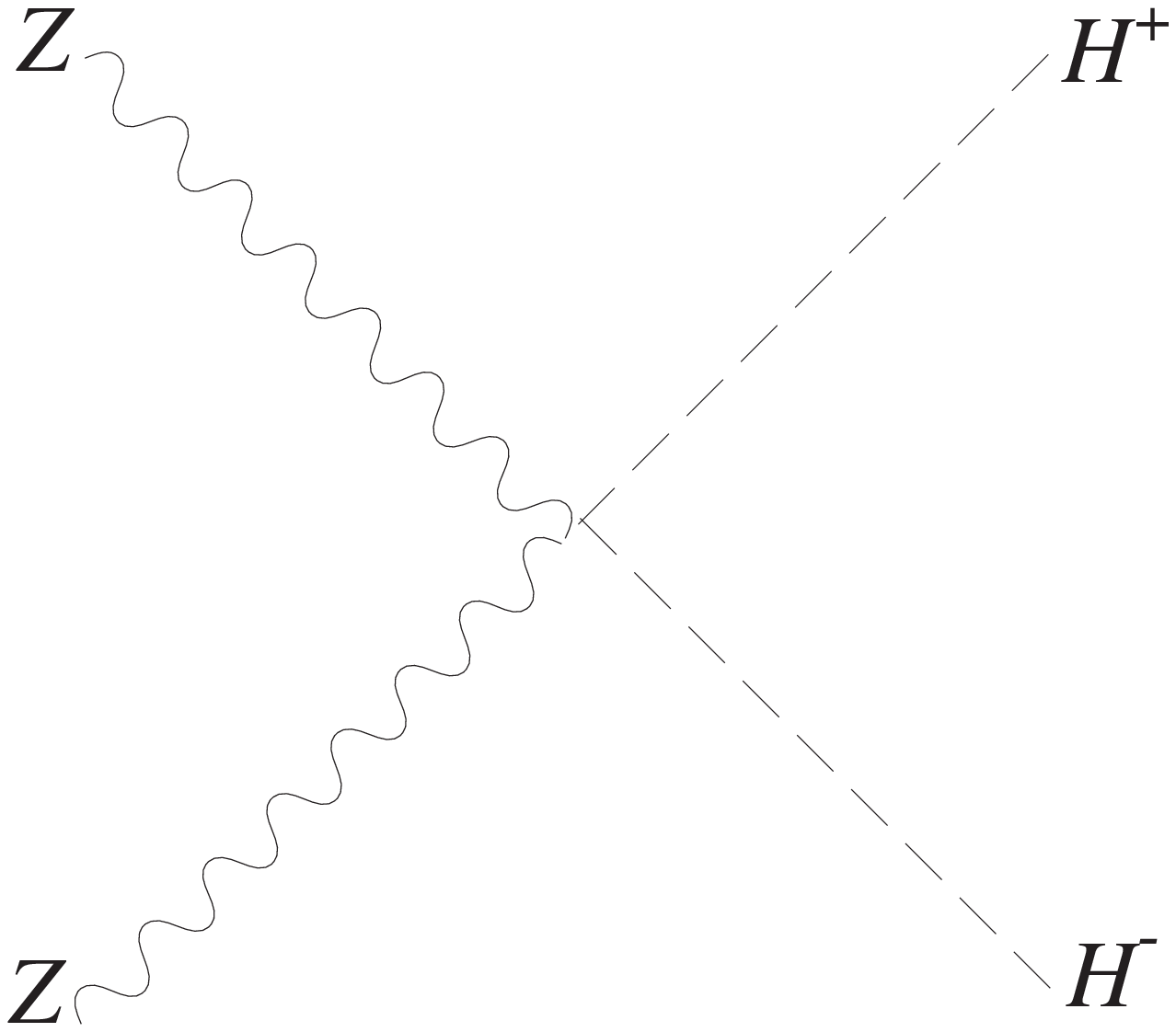}}}
\newcommand{\rfl}{ \mbox{\vcpsboxto{3.5cm}{0cm}{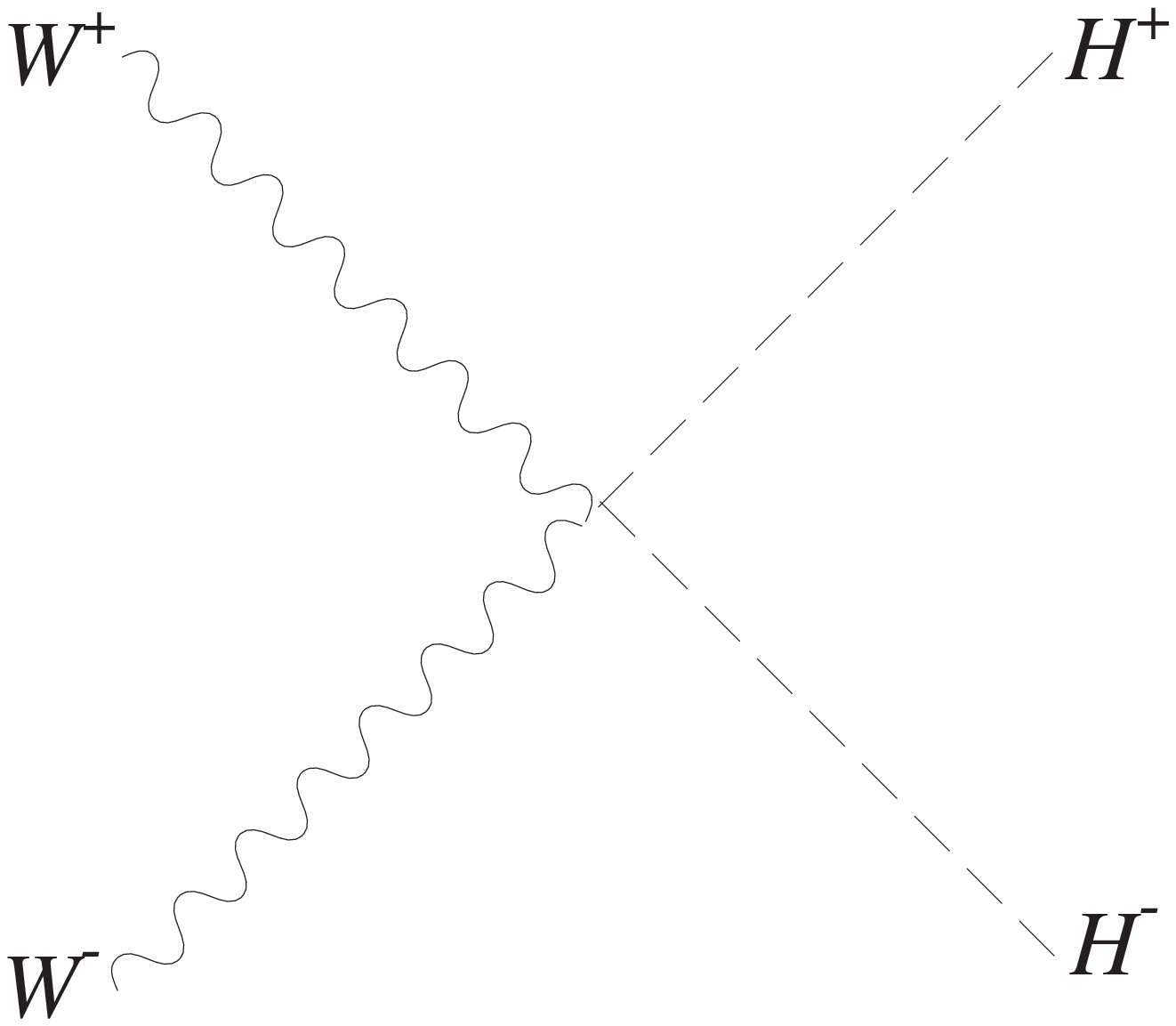}}}
\newcommand{\rfm}{ \mbox{\vcpsboxto{3.5cm}{0cm}{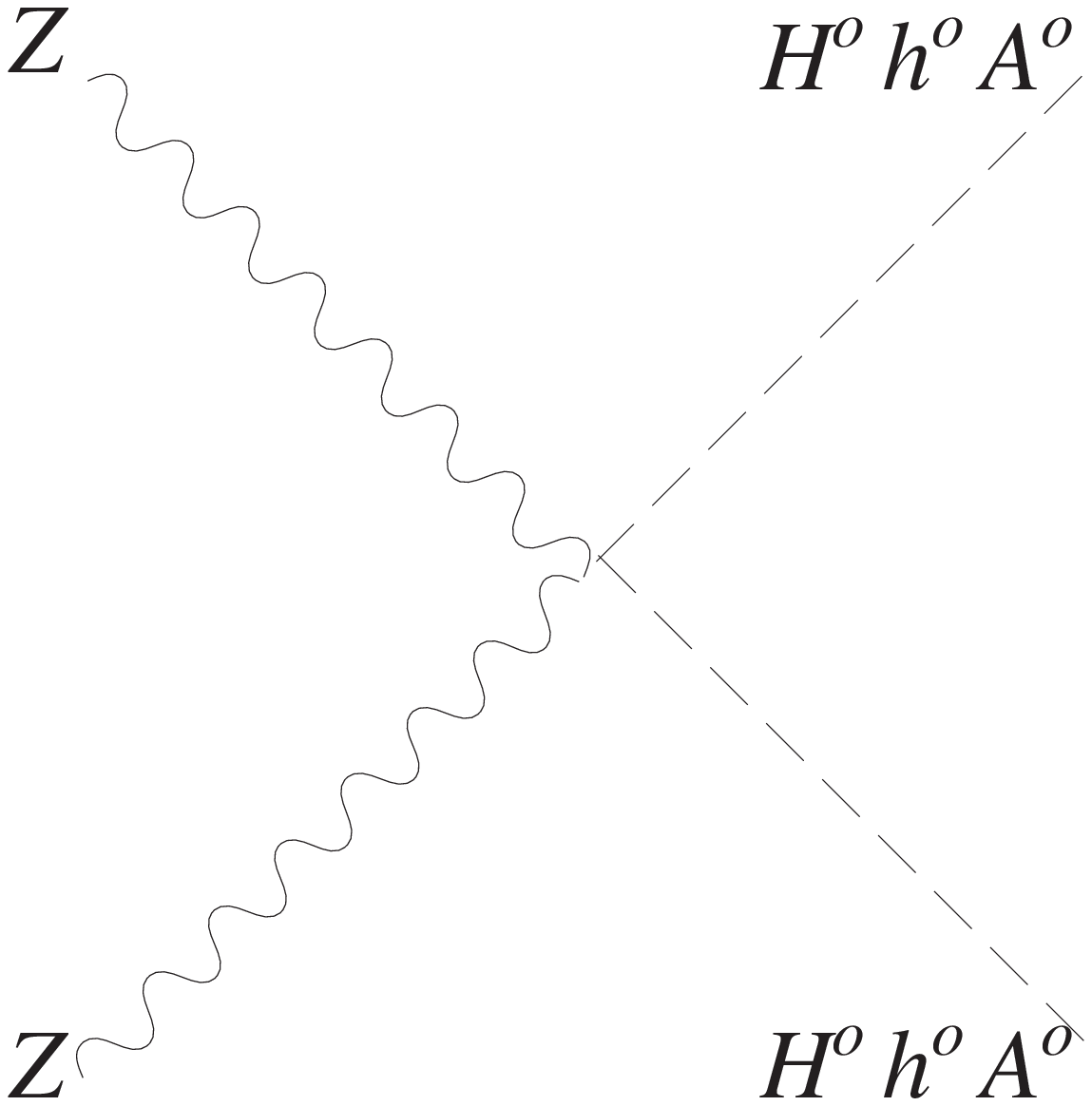}}}
\newcommand{\rfn}{ \mbox{\vcpsboxto{3.5cm}{0cm}{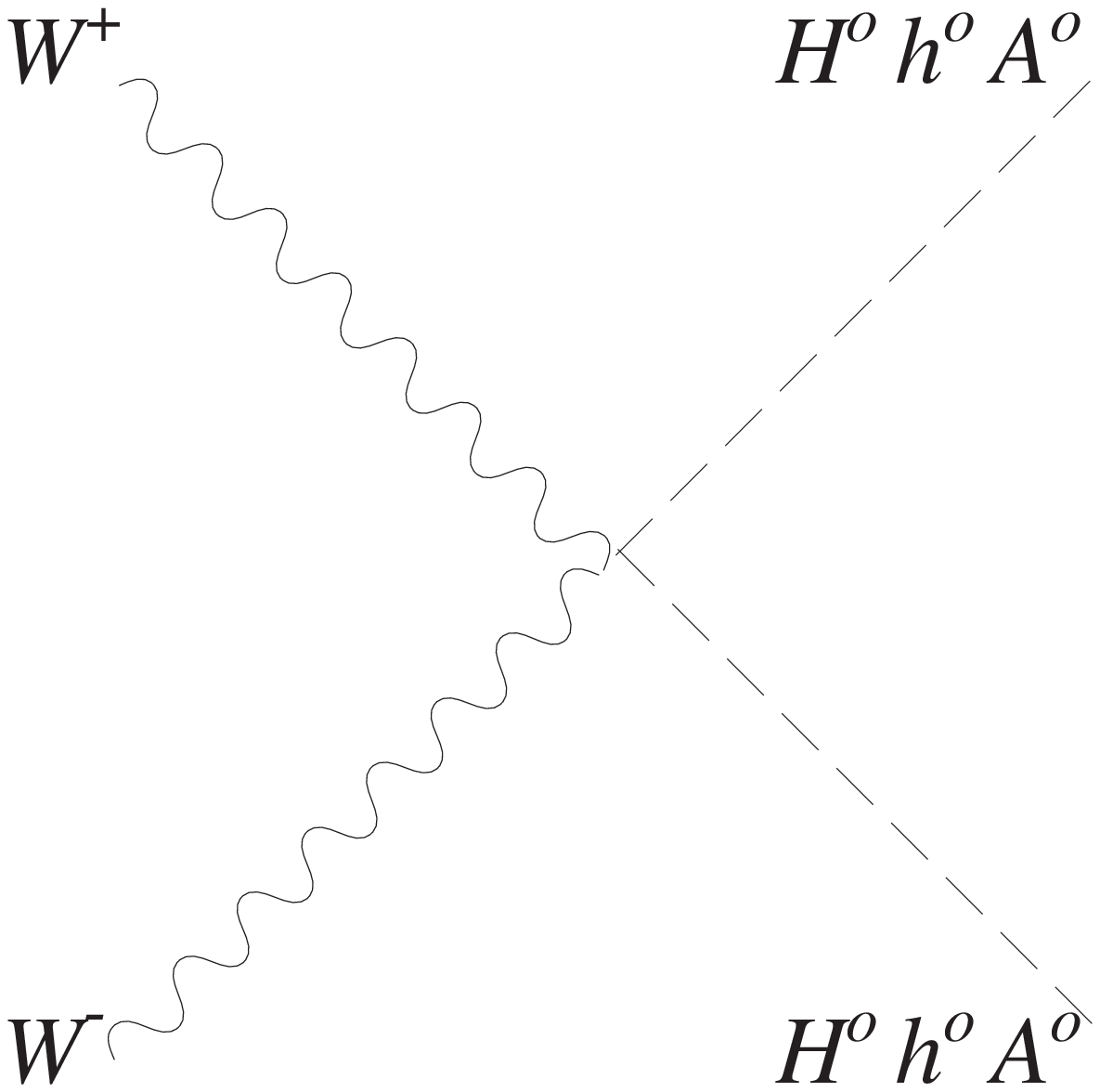}}}

\newcommand{\figutt}{\mbox{\vcpsboxto{3.5cm}{0cm}{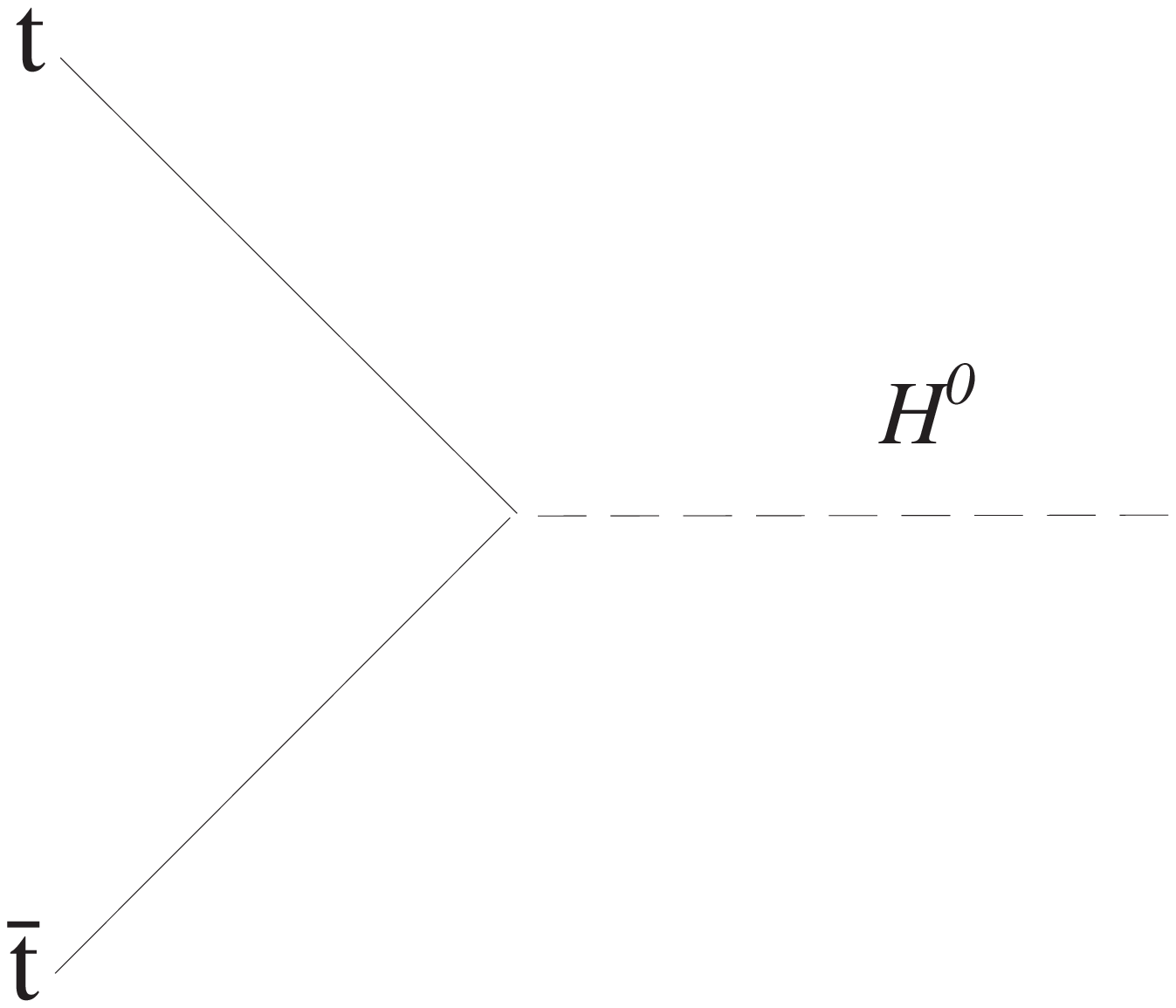}}}
\newcommand{\figubb}{\mbox{\vcpsboxto{3.5cm}{0cm}{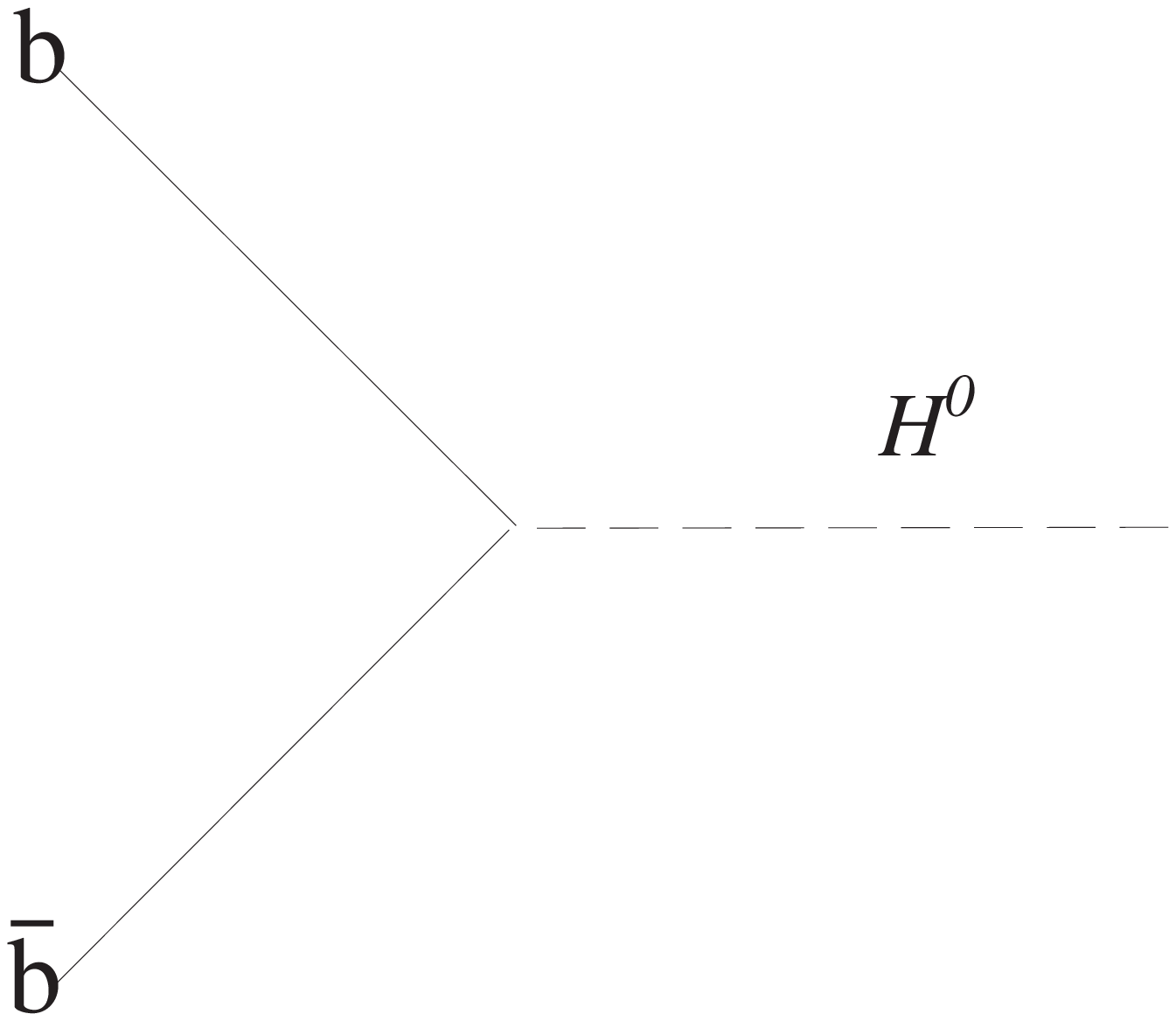}}}
\newcommand{\figltt}{\mbox{\vcpsboxto{3.5cm}{0cm}{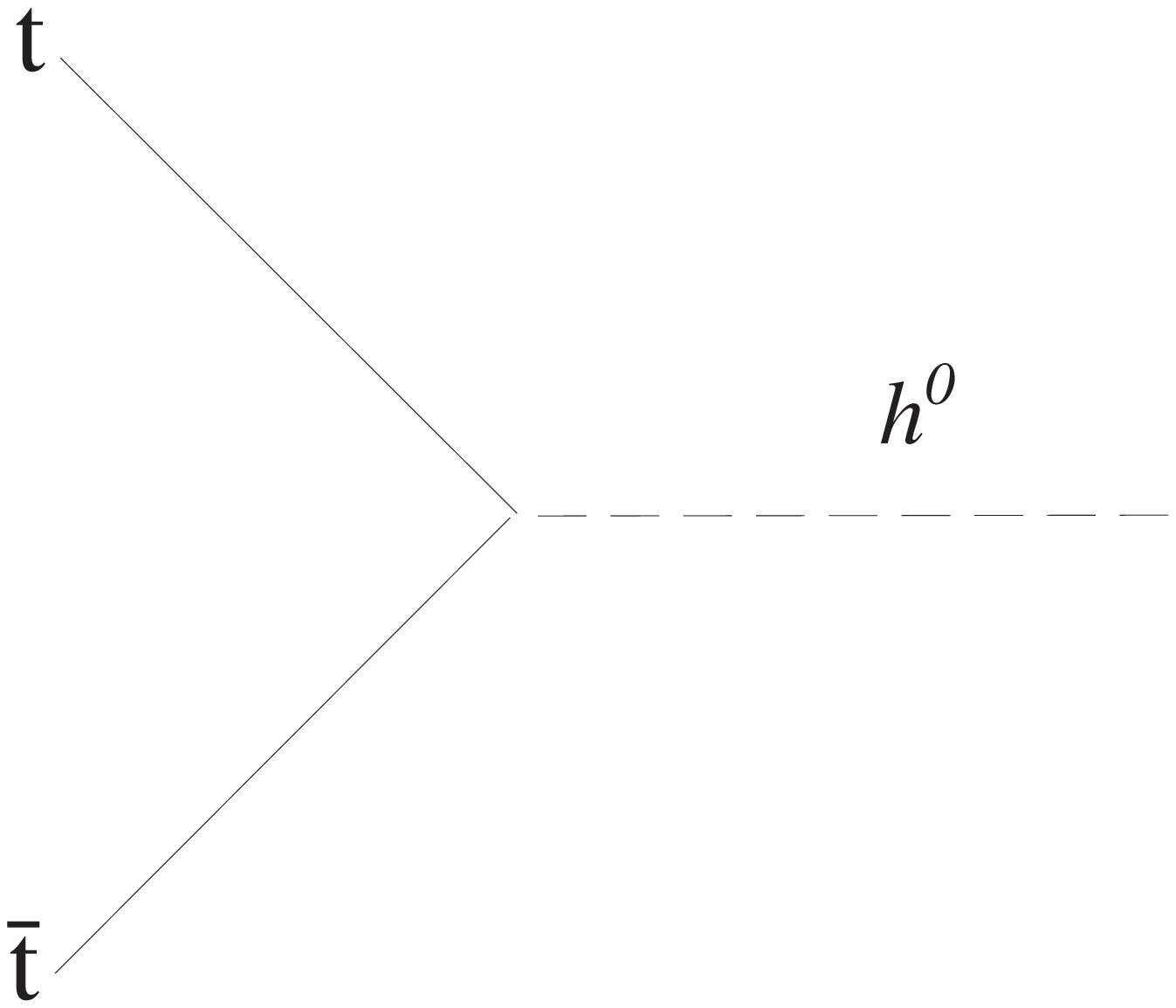}}}
\newcommand{\figlbb}{\mbox{\vcpsboxto{3.5cm}{0cm}{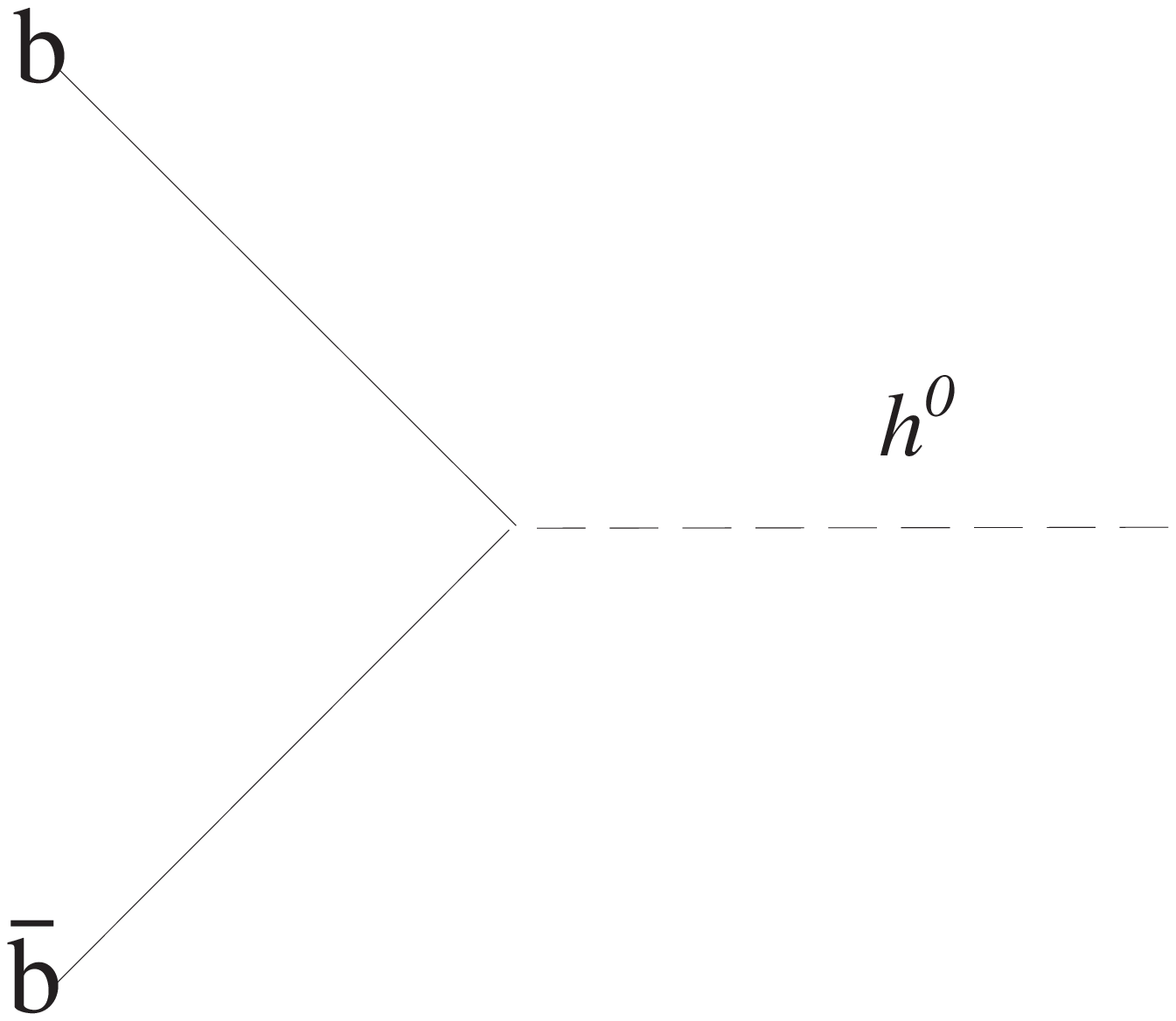}}}
\newcommand{\figatt}{\mbox{\vcpsboxto{3.5cm}{0cm}{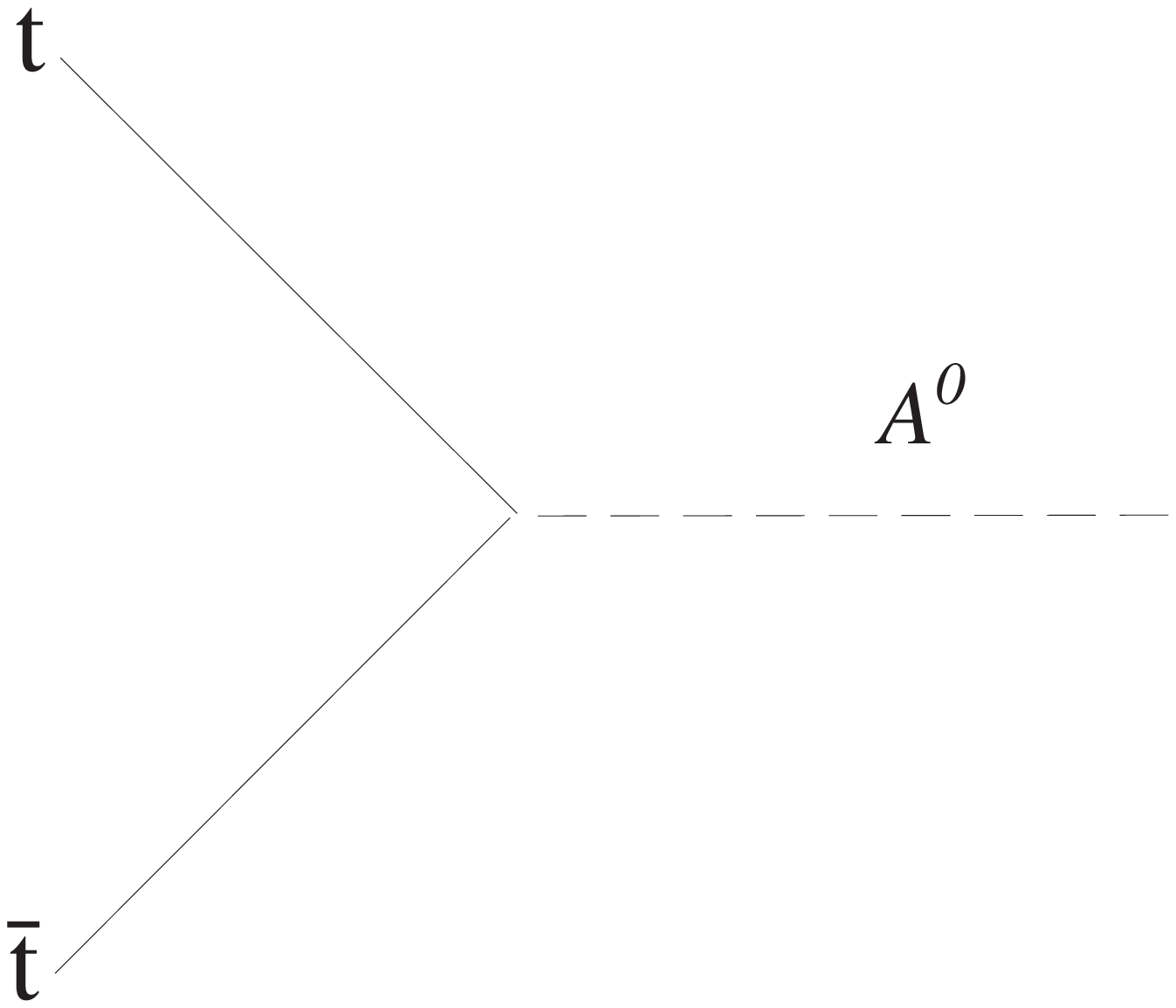}}}
\newcommand{\figabb}{\mbox{\vcpsboxto{3.5cm}{0cm}{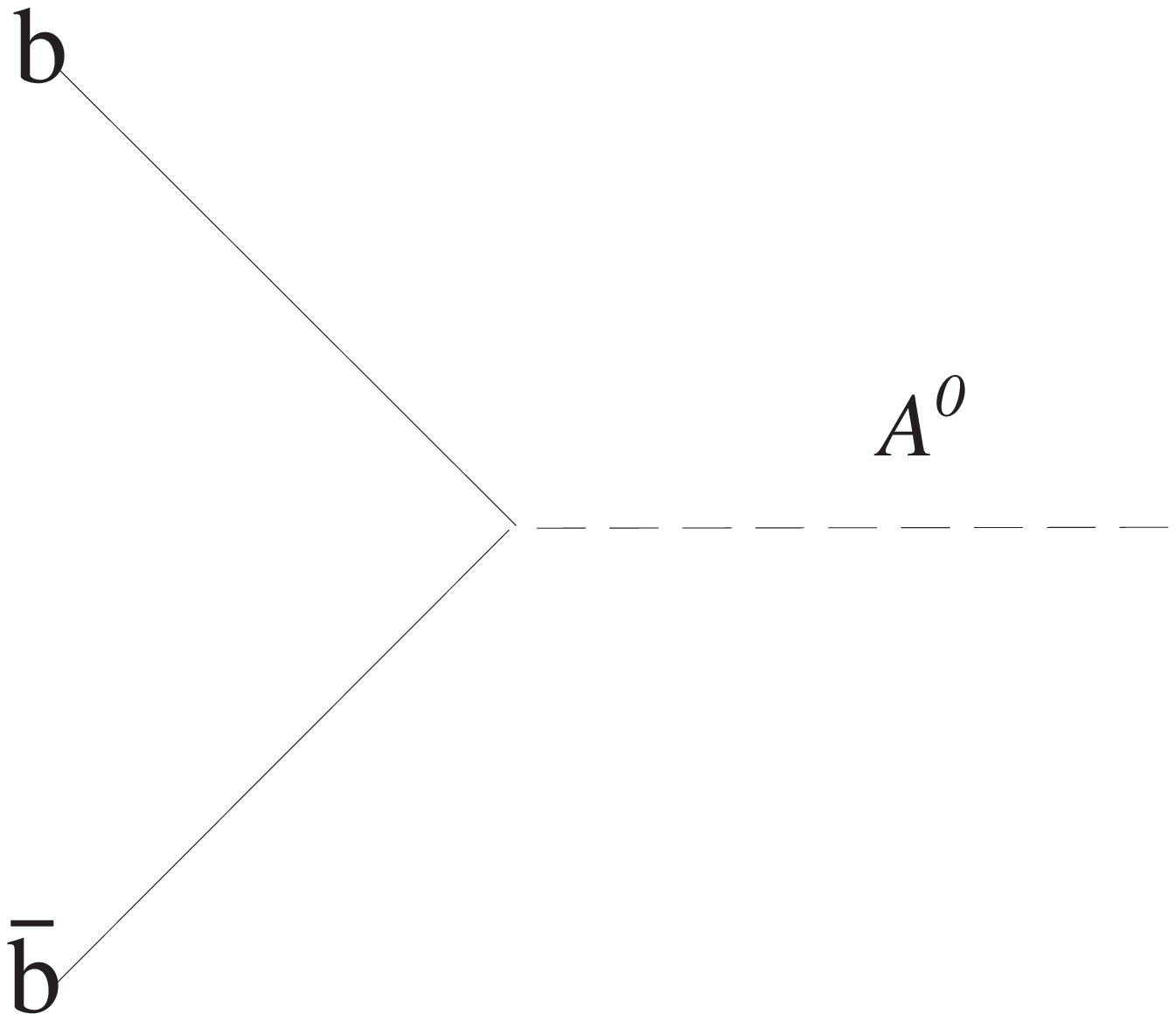}}}
\newcommand{\figptb}{\mbox{\vcpsboxto{3.5cm}{0cm}{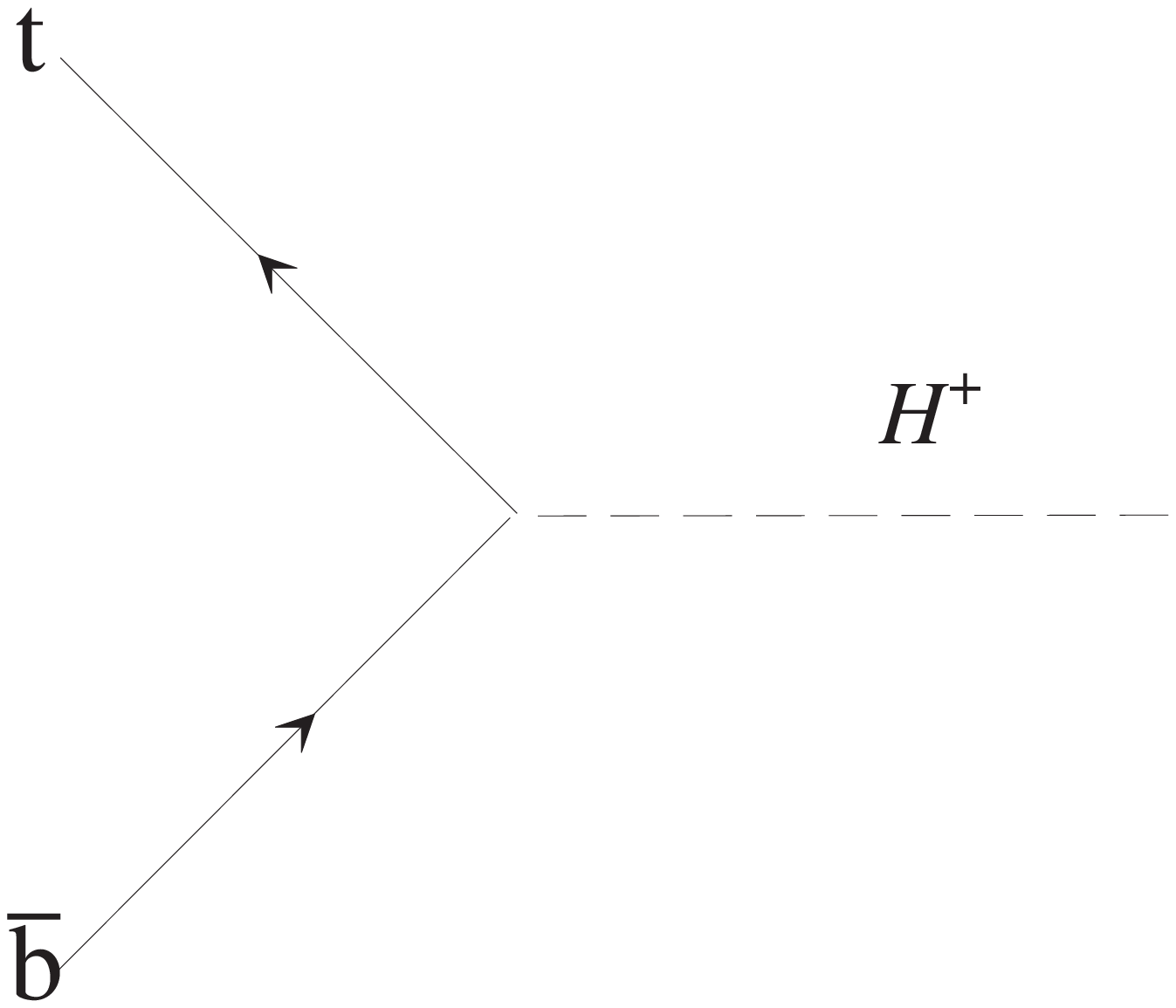}}}
\newcommand{\figmbt}{\mbox{\vcpsboxto{3.5cm}{0cm}{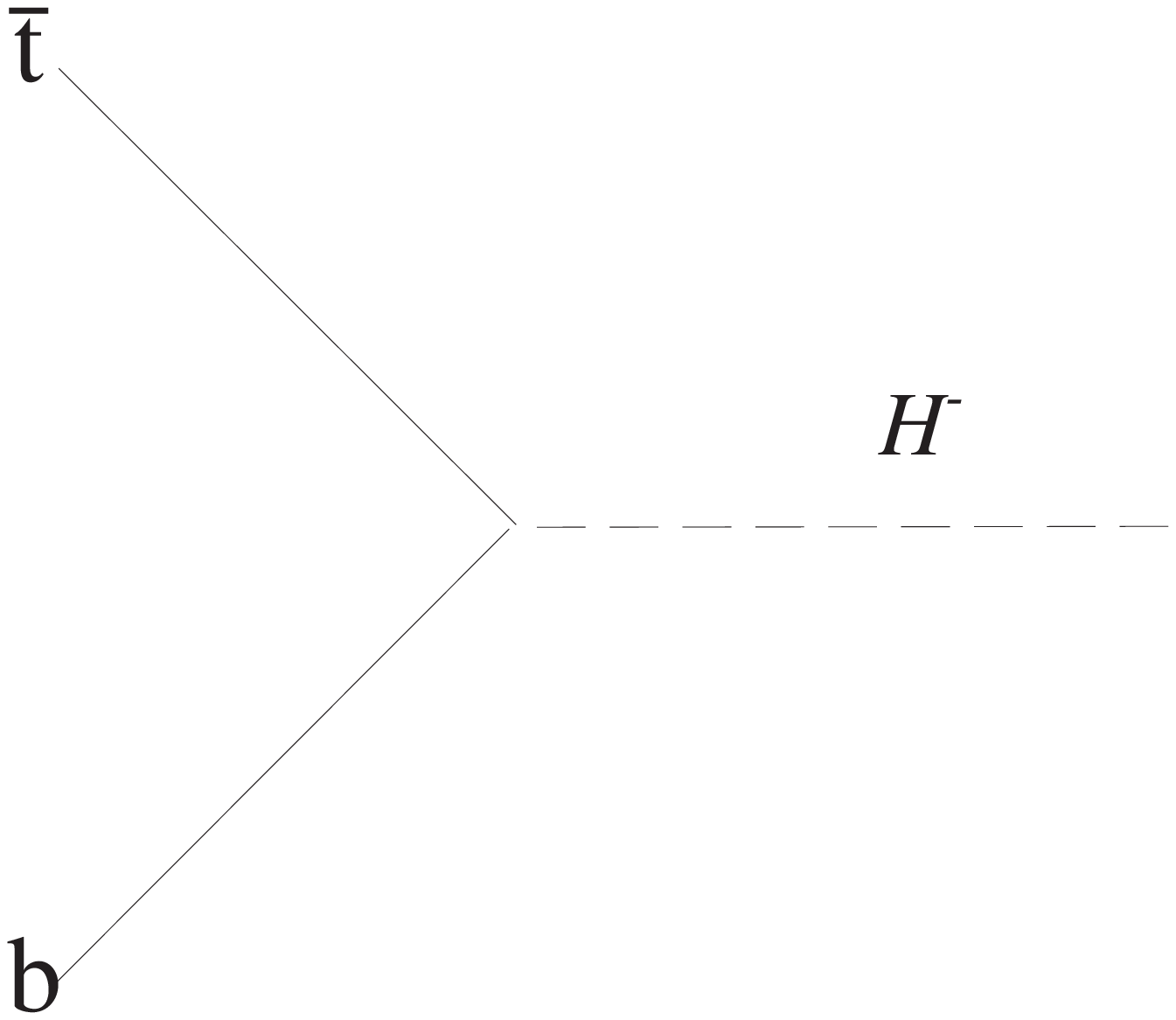}}}

\newcommand{\eqhpmu}
{
\begin{array}[t]{ll}
  \bfrac{ \csc \beta \sec \beta  }{ 2^{5/2} v } &
    \prt{ 2 \MH^2 \prt{  \sin (\alpha + \beta) - \sin (\alpha - 3\beta)
 }
        } + \\
  & \prt{ \MHo^2  \prt{ 3\sin (\alpha + \beta) + \sin (\alpha - 3\beta)
 }
        } \\
\label{eq:hpmu}
\end{array}
}

\newcommand{\eqhpml}
{
\begin{array}[t]{ll}
  \bfrac{ \csc \beta \sec \beta  }{ 2^{5/2} v } &
    \prt{ 2 \MH^2 \prt{ \cos (\alpha + \beta) - \cos (\alpha - 3\beta)
}
        } + \\
  & \prt{ \Mho^2 \prt{3\cos (\alpha +  \beta) + \cos (\alpha - 3\beta)
}
        } \\
\label{eq:hpml}
\end{array}
}

\newcommand{\eqhuuu}
{
\begin{array}[t]{ll}
           \bfrac{ \csc \beta \sec \beta  }{ 2^{7/2} v } &
           \MHo^2 \prt{ 3 \sin \prt{\alpha+\beta}- \sin
\prt{3\alpha+\beta}}\\
\label{eq:huuu}
\end{array}
}

\newcommand{\eqhlll}
{
\begin{array}[t]{ll}
           \bfrac{ \csc \beta \sec \beta  }{ 2^{7/2} v } &
           \Mho^2 \prt{ 3 \cos \prt{\alpha+\beta}+ \cos
\prt{3\alpha+\beta}}\\
\label{eq:hlll}
\end{array}
}

\newcommand{\eqhaau}
{
\begin{array}[t]{ll}
           \bfrac{ \csc \beta \sec \beta  }{ 2^{5/2} v } &
             \prt{ 2 \MAo^2 \prt{ \sin (\alpha +   \beta) -
                                  \sin (\alpha - 3 \beta)  }
                 } + \\
      &      \prt{ \MHo^2 \prt{3\sin (\alpha +  \beta) +
                                \sin (\alpha - 3\beta)  }
                 }\\
\label{eq:haau}
\end{array}
}

\newcommand{\eqhuul}
{
\begin{array}[t]{ll}
           \bfrac{ \csc \beta \sec \beta  }{ 2^{7/2} v } &
           \prt{2\MHo^2 +\Mho^2}\prt{ \sin 2\alpha \sin
\prt{\alpha-\beta}}\\
\label{eq:huul}
\end{array}
}

\newcommand{\eqhllu}
{
\begin{array}[t]{ll}
           \bfrac{ \csc \beta \sec \beta  }{ 2^{7/2} v } &
           \prt{2\Mho^2 +\MHo^2}\prt{ \sin 2\alpha \cos
\prt{\alpha-\beta}}\\
\label{eq:hllu}
\end{array}
}

\newcommand{\eqhaal}
{
\begin{array}[t]{ll}
           \bfrac{ \csc \beta \sec \beta  }{ 2^{5/2} v } &
             \prt{ 2 \MAo^2 \prt{ \cos (\alpha +   \beta) -
                                 \cos (\alpha - 3 \beta)  }
                 } + \\
      &      \prt{ \Mho^2 \prt{3\cos (\alpha +  \beta) +
                                \cos (\alpha - 3\beta)  }
                 }\\
\label{eq:haal}
\end{array}
}

\newcommand{\eqquatrehiggs}
{\be
\begin{array}{lccl}
 \rho_{H^+H^-H^+H^-} & = & \bfrac{\csc^2\beta \sec^2\beta}{64v^2} &
        \left(\;\MHo^2 \prt{\sin(\alpha-3\beta)+3\sin(\alpha+\beta)}^2
+\right.\\
  & & & \;\;\;\left. \Mho^2
\prt{\cos(\alpha-3\beta)+3\cos(\alpha+\beta)}^2
        \;\;\right)\\[1cm]
 \rho_{H^0H^0H^0H^0} & = & \bfrac{\csc^2\beta \sec^2\beta}{256v^2}&
        \left(\;\MHo^2 \prt{\sin(3\alpha-\beta)-3\sin(\alpha+\beta)}^2
+\right.\\
   & & &\;\;\;\left. \Mho^2 \prt{ 2 \sin(2\alpha) \sin(\alpha-\beta)}^2
        \;\;\right)\\[1cm]
 \rho_{A^0A^0A^0A^0} & = & \bfrac{\csc^2\beta \sec^2\beta}{256v^2} &
        \left(\;\MHo^2 \prt{\sin(\alpha-3\beta)+3\sin(\alpha+\beta)}^2
+\right.\\
   & & &\;\;\;\left. \Mho^2
\prt{\cos(\alpha-3\beta)+3\cos(\alpha+\beta)}^2
        \;\;\right)\\[1cm]
 \rho_{h^0h^0h^0h^0} & = & \bfrac{\csc^2\beta \sec^2\beta}{256v^2} &
         \left(\;\MHo^2 \prt{ 2 \sin(2\alpha) \cos(\alpha-\beta)}^2
+\right.   \\
   & & &\;\;\;\left. \Mho^2
\prt{\cos(3\alpha-\beta)+3\cos(\alpha+\beta)}^2
        \;\;\right)\\
\end{array}
\ee}

\newcommand{\figpmcinccents}{
\bfig
   \centerline
   {
     \begin{tabular}{r@{\hspace{-0.05cm}}c@{\hspace{1cm}}r@{\hspace{-0.05cm}}c}
       \multicolumn{2}{c}{
       $\scriptstyle\MHo\!=100GeV, \;\;\Mho\!=80GeV, \;\;\MAo\!=100GeV$} &
       \multicolumn{2}{c}{
       $\scriptstyle\MHo\!=100GeV, \;\;\Mho\!=80GeV, \;\;\MAo\!=300GeV$} \\
       $\alpha$ &\mbox{ \pshtincr=-3.8cm\psyoffset=-1.9cm
                        \vcpsboxto{7cm}{0cm}{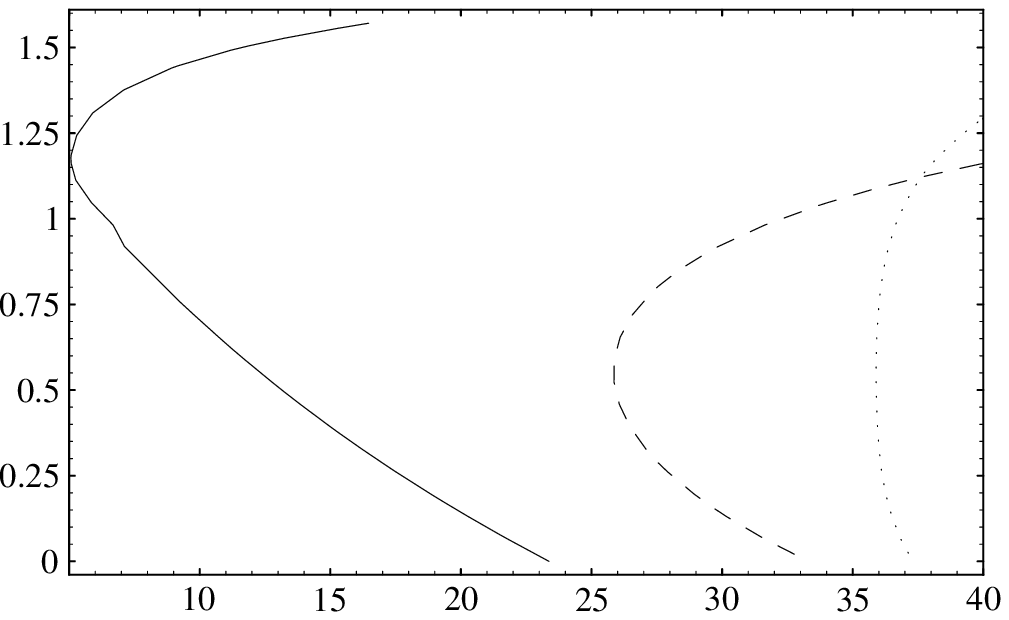}  }&
       $\alpha$ &\mbox{ \pshtincr=-3.8cm\psyoffset=-1.9cm
                        \vcpsboxto{7cm}{0cm}{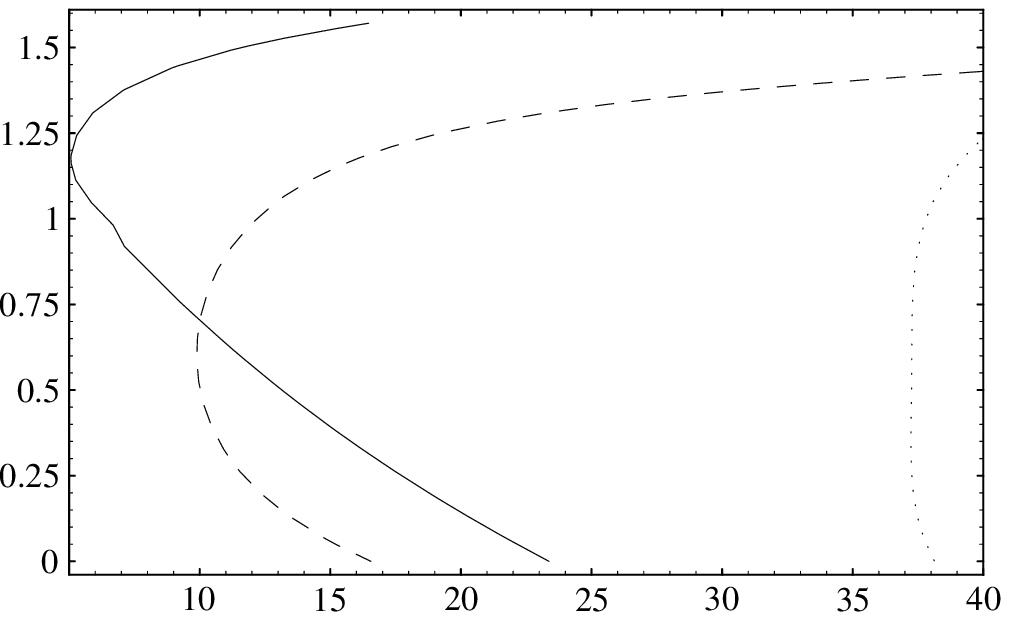}  }\\[-0.2cm]
                & $\scriptstyle\tan\beta$ &
                & $\scriptstyle\tan\beta$ \\[1cm]
       \multicolumn{2}{c}{
       $\scriptstyle\MHo\!=120GeV, \;\;\Mho\!=100GeV, \;\;\MAo\!=100GeV$} &
       \multicolumn{2}{c}{
       $\scriptstyle\MHo\!=120GeV, \;\;\Mho\!=100GeV, \;\;\MAo\!=300GeV$} \\
       $\alpha$ &\mbox{ \pshtincr=-3.8cm\psyoffset=-1.9cm
                        \vcpsboxto{7cm}{0cm}{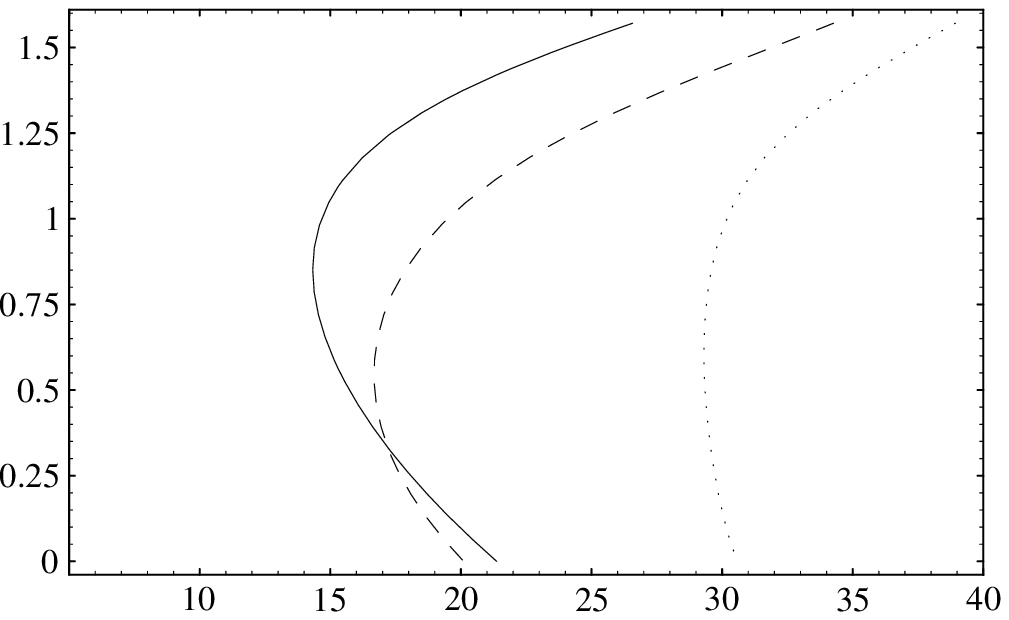}  }&
       $\alpha$ &\mbox{ \pshtincr=-3.8cm\psyoffset=-1.9cm
                        \vcpsboxto{7cm}{0cm}{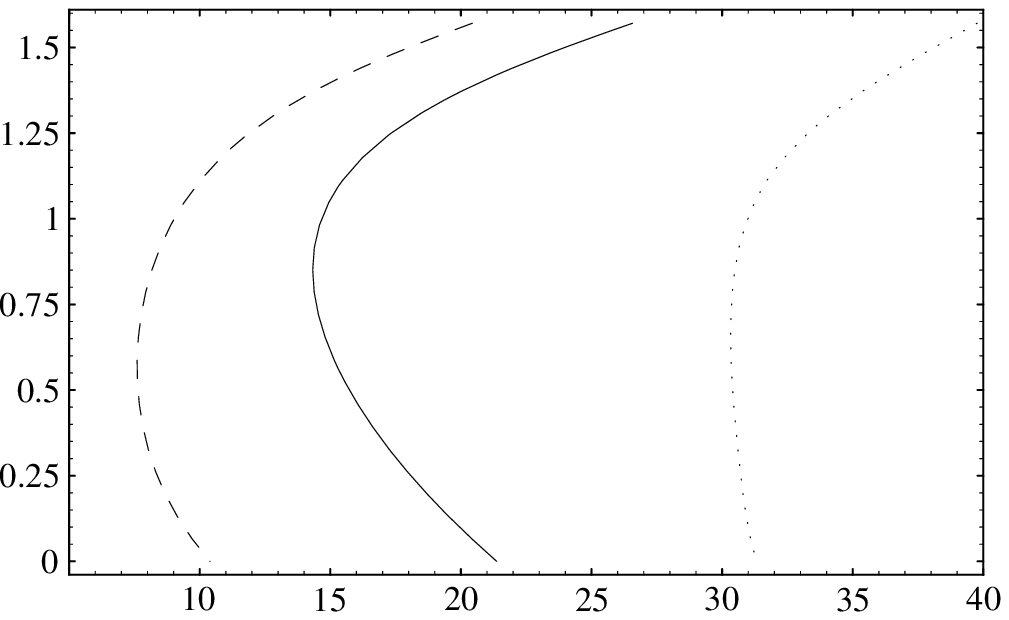}  }\\[-0.2cm]
                & $\scriptstyle\tan\beta$ &
                & $\scriptstyle\tan\beta$
     \end{tabular}
   }
   {
   \hspace*{\fill}
   \parbox{13cm}{ \caption{\em{
       Allowed (shaded) areas in the $\alpha$ and $\tan\!\beta$
       plane for $\MH=500GeV$ and various choices of neutral
       Higgs boson masses.
   }}\label{fig:pm500}}
   \hspace*{\fill}
   }
\efig
}

\newcommand{\figpmquatrecents}{
\bfig
   \centerline
   {
     \begin{tabular}{r@{\hspace{-0.05cm}}c@{\hspace{1cm}}r@{\hspace{-0.05cm}}c}
       \multicolumn{2}{c}{
       $\scriptstyle\MHo\!=100GeV, \;\;\Mho\!=80GeV, \;\;\MAo\!=100GeV$} &
       \multicolumn{2}{c}{
       $\scriptstyle\MHo\!=100GeV, \;\;\Mho\!=80GeV, \;\;\MAo\!=300GeV$} \\
       $\alpha$ &\mbox{ \pshtincr=-3.8cm\psyoffset=-1.9cm
                        \vcpsboxto{7cm}{0cm}{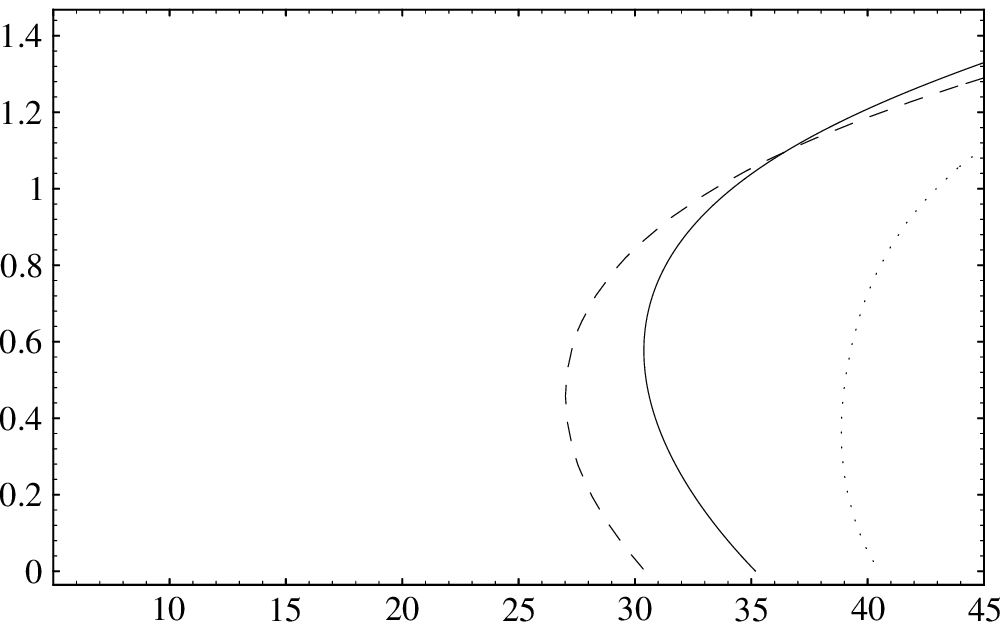}  }&
       $\alpha$ &\mbox{ \pshtincr=-3.8cm\psyoffset=-1.9cm
                        \vcpsboxto{7cm}{0cm}{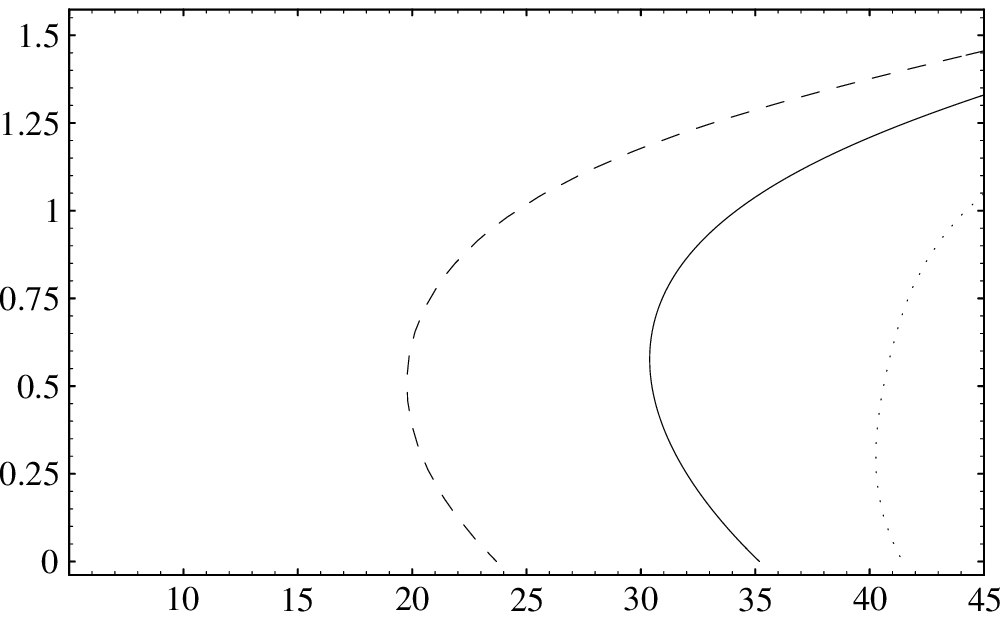}  }\\[-0.2cm]
                & $\scriptstyle\tan\beta$ &
                & $\scriptstyle\tan\beta$ \\[1cm]
       \multicolumn{2}{c}{
       $\scriptstyle\MHo\!=120GeV, \;\;\Mho\!=100GeV, \;\;\MAo\!=100GeV$} &
       \multicolumn{2}{c}{
       $\scriptstyle\MHo\!=120GeV, \;\;\Mho\!=100GeV, \;\;\MAo\!=300GeV$} \\
       $\alpha$ &\mbox{ \pshtincr=-3.8cm\psyoffset=-1.9cm
                        \vcpsboxto{7cm}{0cm}{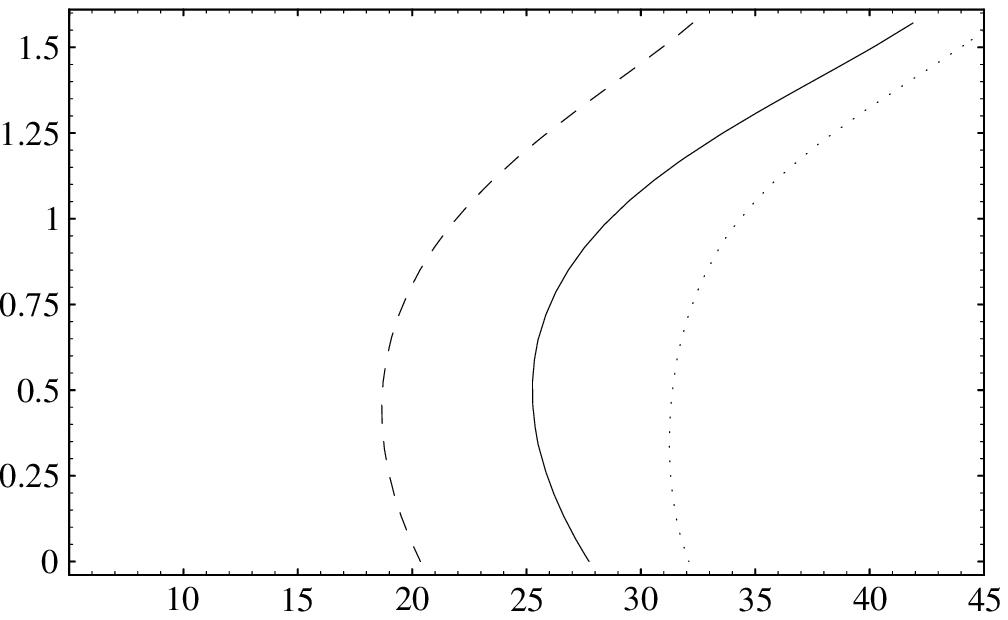}  }&
       $\alpha$ &\mbox{ \pshtincr=-3.8cm\psyoffset=-1.9cm
                        \vcpsboxto{7cm}{0cm}{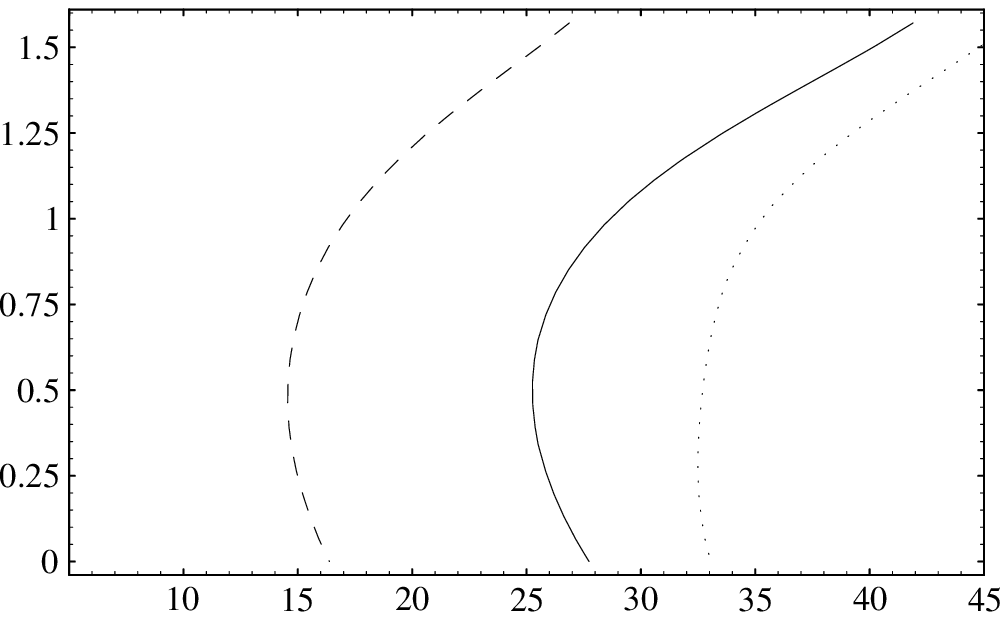}  }\\[-0.2cm]
                & $\scriptstyle\tan\beta$ &
                & $\scriptstyle\tan\beta$
     \end{tabular}
   }
   {
   \hspace*{\fill}
   \parbox{13cm}{ \caption{\em{
       Allowed (shaded) areas in the $\alpha$ and $\tan\!\beta$
       plane for $\MH=400GeV$ and various choices of neutral
       Higgs boson masses.
   }}\label{fig:pm400}}
   \hspace*{\fill}
   }
\efig
}

\newcommand{\figpmtrescents}{
\bfig
   \centerline
   {
     \begin{tabular}{r@{\hspace{-0.05cm}}c@{\hspace{1cm}}r@{\hspace{-0.05cm}}c}
       \multicolumn{2}{c}{
       $\scriptstyle\MHo\!=100GeV, \;\;\Mho\!=80GeV, \;\;\MAo\!=100GeV$} &
       \multicolumn{2}{c}{
       $\scriptstyle\MHo\!=100GeV, \;\;\Mho\!=80GeV, \;\;\MAo\!=300GeV$} \\
       $\alpha$ &\mbox{ \pshtincr=-3.8cm\psyoffset=-1.9cm
                        \vcpsboxto{7cm}{0cm}{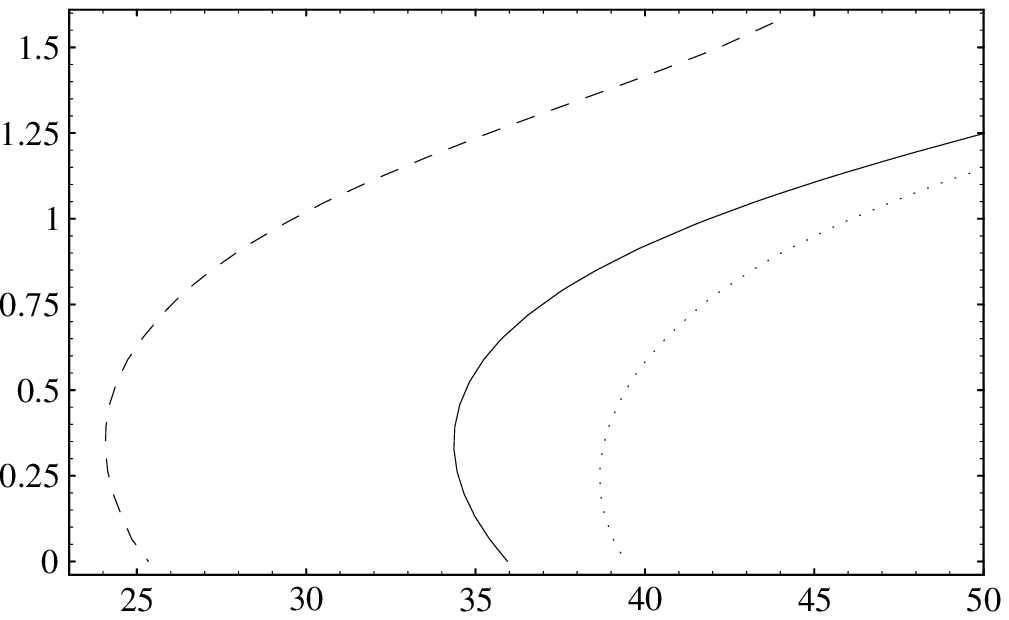}  }&
       $\alpha$ &\mbox{ \pshtincr=-3.8cm\psyoffset=-1.9cm
                        \vcpsboxto{7cm}{0cm}{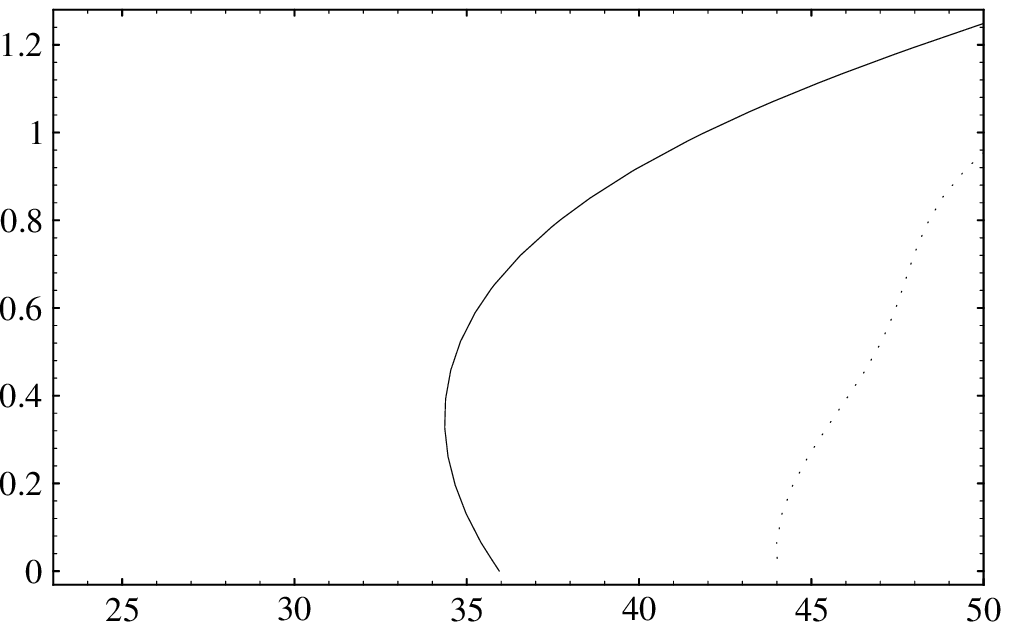}  }\\[-0.2cm]
                & $\scriptstyle\tan\beta$ &
                & $\scriptstyle\tan\beta$ \\[1cm]
       \multicolumn{2}{c}{
       $\scriptstyle\MHo\!=120GeV, \;\;\Mho\!=100GeV, \;\;\MAo\!=100GeV$} &
       \multicolumn{2}{c}{
       $\scriptstyle\MHo\!=120GeV, \;\;\Mho\!=100GeV, \;\;\MAo\!=300GeV$} \\
       $\alpha$ &\mbox{ \pshtincr=-3.8cm\psyoffset=-1.9cm
                        \vcpsboxto{7cm}{0cm}{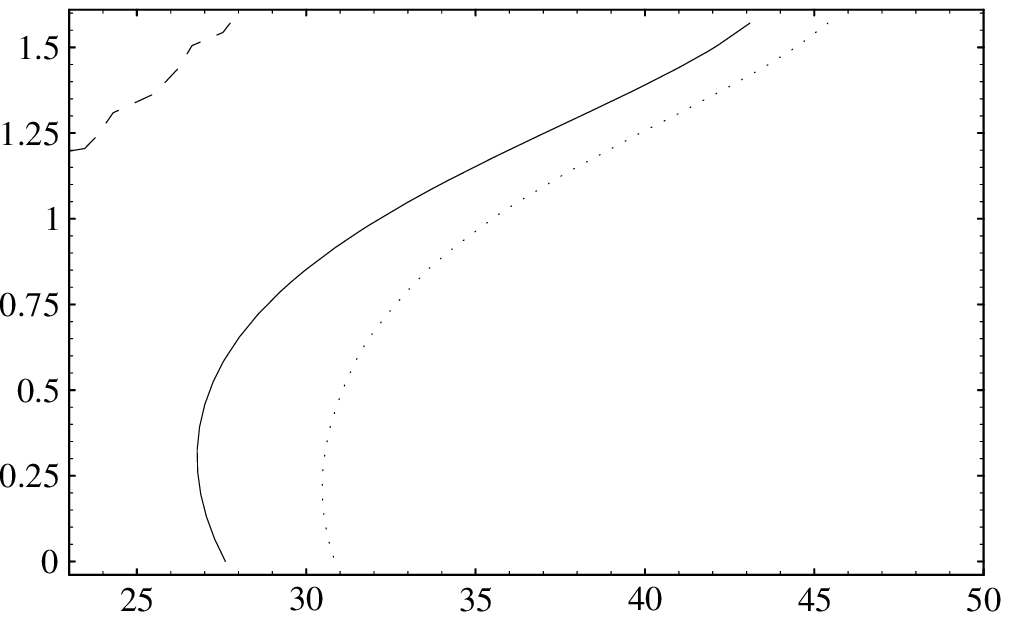}  }&
       $\alpha$ &\mbox{ \pshtincr=-3.8cm\psyoffset=-1.9cm
                        \vcpsboxto{7cm}{0cm}{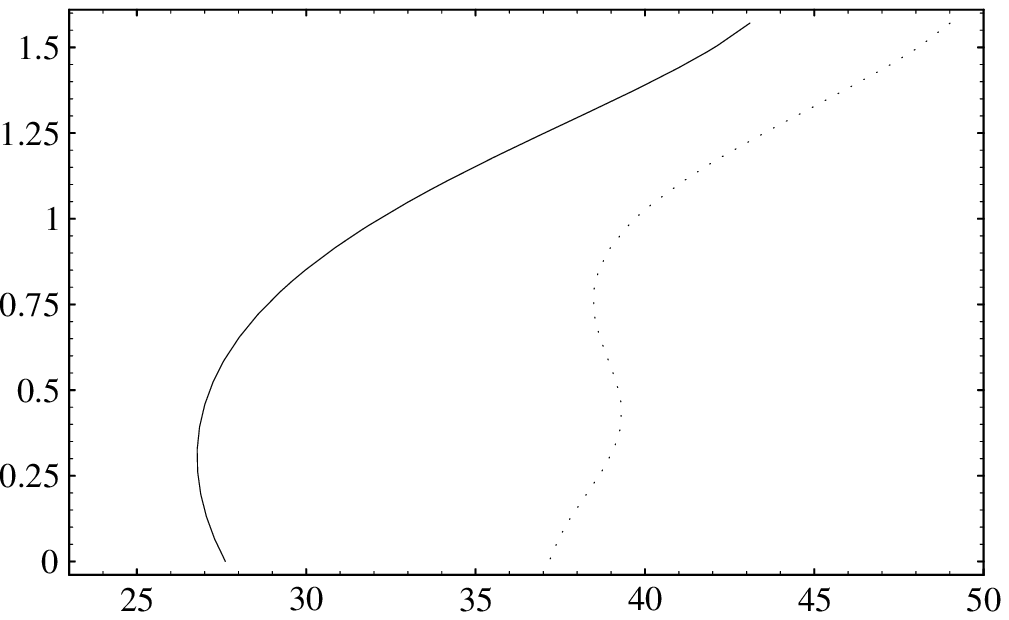}  }\\[-0.2cm]
                & $\scriptstyle\tan\beta$ &
                & $\scriptstyle\tan\beta$
     \end{tabular}
   }
   {
   \hspace*{\fill}
   \parbox{13cm}{ \caption{\em{
       Allowed (shaded) areas in the $\alpha$ and $\tan\!\beta$
       plane for $\MH=300GeV$ and various choices of neutral
       Higgs boson masses.
   }}\label{fig:pm300}}
   \hspace*{\fill}
   }
\efig
}

\newcommand{\figaaquatrecents}{
\bfig
   \centerline
   {
     \begin{tabular}{r@{\hspace{-0.05cm}}c@{\hspace{1cm}}r@{\hspace{-0.05cm}}c}
       \multicolumn{2}{c}{
       $\scriptstyle\MHo\!=100GeV, \;\;\Mho\!=80GeV, \;\;\MH\!=100GeV$} &
       \multicolumn{2}{c}{
       $\scriptstyle\MHo\!=100GeV, \;\;\Mho\!=80GeV, \;\;\MH\!=400GeV$} \\
       $\alpha$ &\mbox{ \pshtincr=-3.8cm\psyoffset=-1.9cm
                        \vcpsboxto{7cm}{0cm}{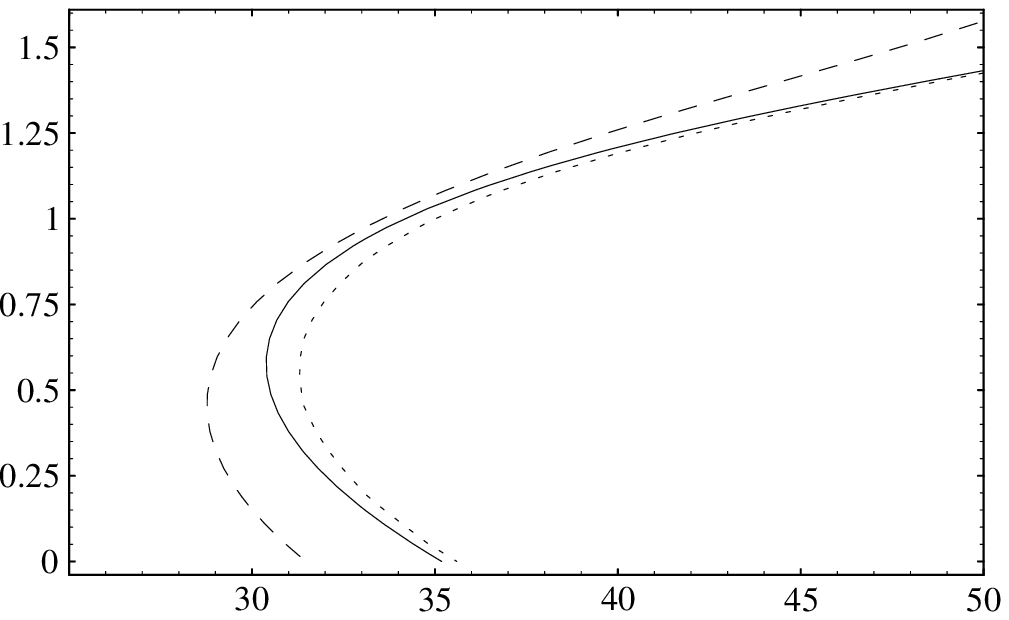}  }&
       $\alpha$ &\mbox{ \pshtincr=-3.8cm\psyoffset=-1.9cm
                        \vcpsboxto{7cm}{0cm}{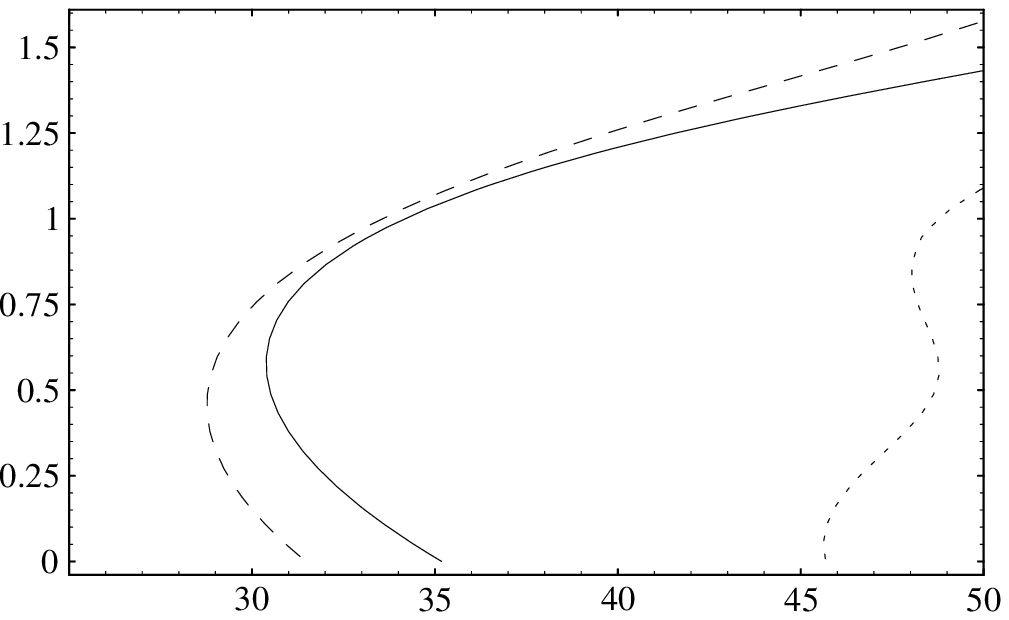}  }\\[-0.2cm]
                & $\scriptstyle\tan\beta$ &
                & $\scriptstyle\tan\beta$ \\[1cm]
       \multicolumn{2}{c}{
       $\scriptstyle\MHo\!=120GeV, \;\;\Mho\!=100GeV, \;\;\MH\!=100GeV$} &
       \multicolumn{2}{c}{
       $\scriptstyle\MHo\!=120GeV, \;\;\Mho\!=100GeV, \;\;\MH\!=400GeV$} \\
       $\alpha$ &\mbox{ \pshtincr=-3.8cm\psyoffset=-1.9cm
                        \vcpsboxto{7cm}{0cm}{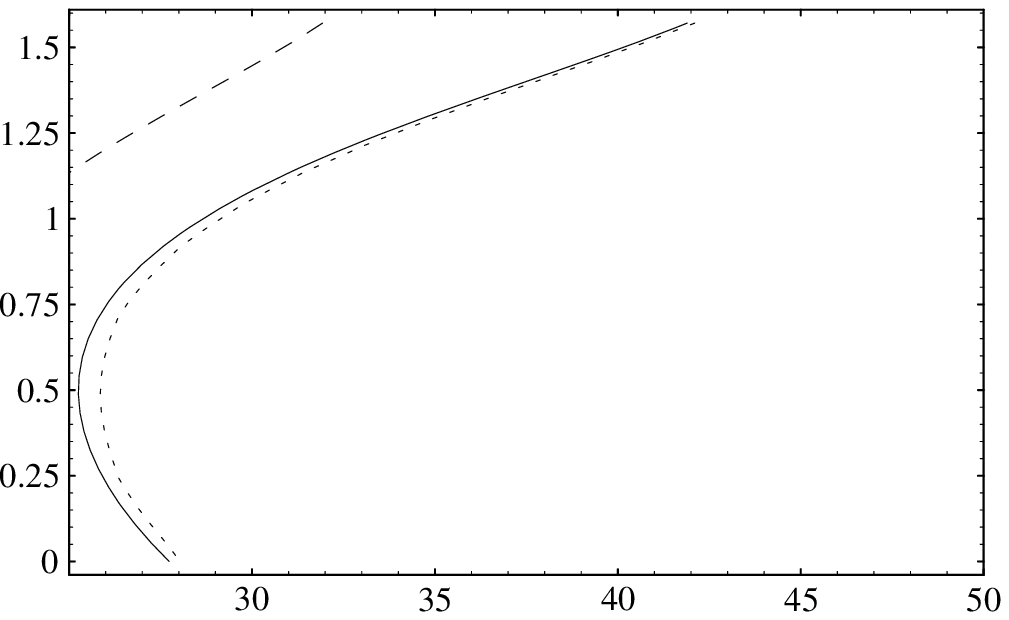}  }&
       $\alpha$ &\mbox{ \pshtincr=-3.8cm\psyoffset=-1.9cm
                        \vcpsboxto{7cm}{0cm}{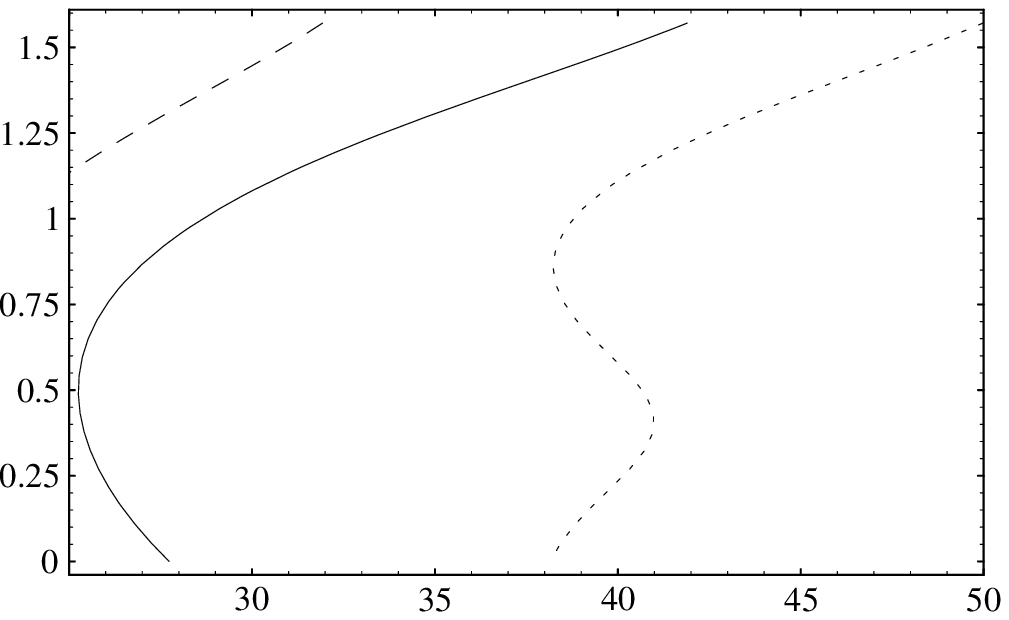}  }\\[-0.2cm]
                & $\scriptstyle\tan\beta$ &
                & $\scriptstyle\tan\beta$
     \end{tabular}
   }
   {
   \hspace*{\fill}
   \parbox{13cm}{ \caption{\em{
       Allowed (shaded) areas in the $\alpha$ and $\tan\!\beta$
       plane for $\MAo=400GeV$ and various choices of neutral
       Higgs boson masses.
   }}\label{fig:aa400}}
   \hspace*{\fill}
   }
\efig
}

\newcommand{\figaacinccents}{
\bfig
   \centerline
   {
     \begin{tabular}{r@{\hspace{-0.05cm}}c@{\hspace{1cm}}r@{\hspace{-0.05cm}}c}
       \multicolumn{2}{c}{
       $\scriptstyle\MHo\!=100GeV, \;\;\Mho\!=80GeV, \;\;\MH\!=500GeV$} &
       \multicolumn{2}{c}{
       $\scriptstyle\MHo\!=120GeV, \;\;\Mho\!=100GeV, \;\;\MH\!=500GeV$} \\
       $\alpha$ &\mbox{ \pshtincr=-3.8cm\psyoffset=-1.9cm
                        \vcpsboxto{7cm}{0cm}{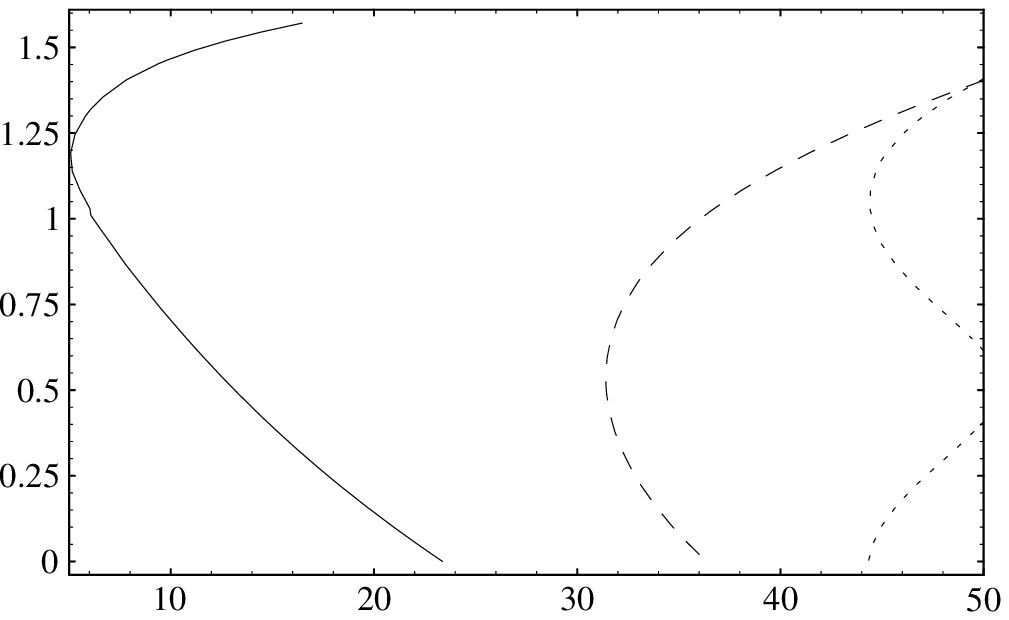}  }&
       $\alpha$ &\mbox{ \pshtincr=-3.8cm\psyoffset=-1.9cm
                        \vcpsboxto{7cm}{0cm}{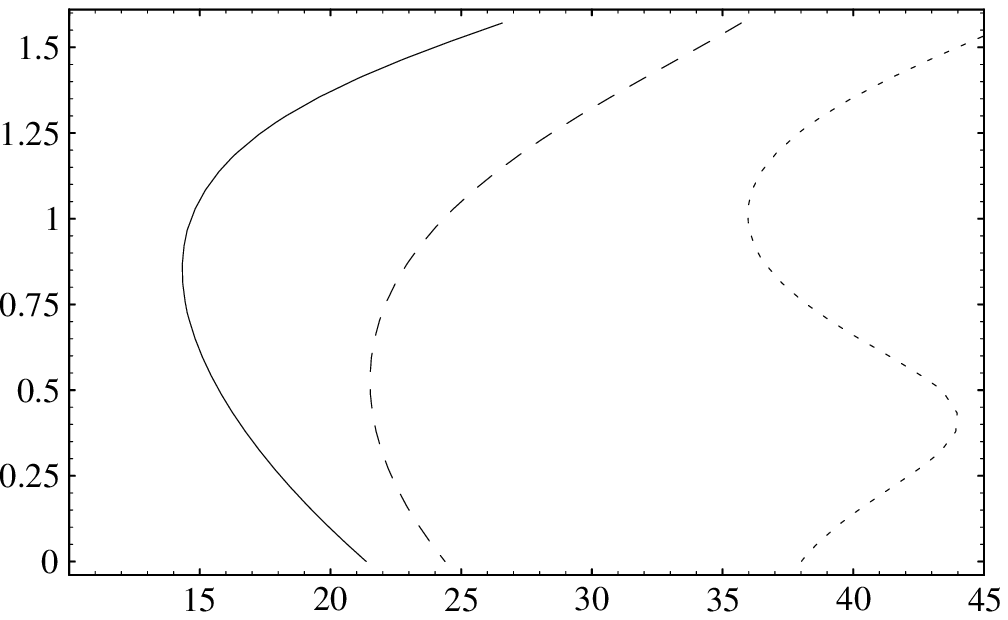}  }\\[-0.2cm]
                & $\scriptstyle\tan\beta$ &
                & $\scriptstyle\tan\beta$
     \end{tabular}
   }
   {
   \hspace*{\fill}
   \parbox{13cm}{ \caption{\em{
       Allowed (shaded) areas in the $\alpha$ and $\tan\!\beta$
       plane for $\MAo=500GeV$ and various choices of neutral
       Higgs boson masses.
   }}\label{fig:aa500}}
   \hspace*{\fill}
   }
\efig
}

\begin{document}

\thispagestyle{empty}

\begin{flushright}
\parbox{3.5cm}{ UAB-FT-364 \\ hep-ph/9505287 }
\end{flushright}

\vspace{3cm}

\begin{center}

{\large \bf

       Bound states of scalar bosons \\
       in extensions of the Standard Model
}

\vspace{0.8cm}

       J. Clua\footnote{e-mail address:{\tt clua@ifae.es}} and
       J. A. Grifols\footnote{e-mail address:{\tt grifols@ifae.es}}
       \vspace{0.3cm}\\
       IFAE. Grup de F\'{\i}sica Te\`orica.\\
       Universitat Aut\`onoma de Barcelona\\
       Bellaterra. Catalonia (Spain)

\end{center}

\vspace{2cm}

\begin{abstract}

We explore systematically, in a general two Higgs doublet model,
the possibility
that bound systems of scalar bosons do exist. We find a wide region
of parameter space  in the scalar potential for which S-wave bound
states of Higgs bosons do indeed exist. On the contrary we show that the
Minimal Supersimmetric Standard Model does not admit such bound systems.
\end{abstract}


\newpage

\section{Introduction}

Progress in a deeper understanding of the Standard Model must probably
follow from an elucidation of the r$\hat o$le and inner workings of
the Higgs boson sector
in the theory. In this respect, of course, the experimental discovery
of the Higgs scalar is crucial to the whole modern particle physics
paradigm.
This is therefore an extremely valuable experimental endevour that should
be pursued in present and future machines. Still, on the theoretical side
one would like to give answers to questions like: is the Higgs boson
a fundamental particle, as quarks and leptons are or, else, is it
a composite object? will its presence signal the existence of a new
strong interacting regime of the weak interactions or, on the contrary
will we get clues to deeper and most embrancing theoretical schemes,
like Grand Unification or Supersymmetry?

Surely, the minimal standard model  requires only one scalar boson
doublet
and it might well be that Nature makes use only of these minimal
degrees of freedom. However, economy of Principles seems to be the
guiding rule rather then economy of structures. Indeed, there is no
present understanding for, e.g., the repetition of fermions of quarks
and leptons. Furthermore, any extension of the standard gauge model
paradigm that tries to encompass larger domains of reality is bound to
contain a larger set of higgs scalar bosons. The Minimal Supersymmetric
Standard  Model (MSSM), for
instance, requires two higgs boson doublets for it to be
phenomenologically sound.

It seems therefore that scalar bosons or effective fields
that, at the Fermi scale
play the r$\hat o$le of the Higgs boson degrees of freedom, are
unavoidable in any sensible construction of a consistent paradigm of
the fundamental physical laws. A rich phenomenology is therefore likely
to develop involving the scalar sector of Physics Beyond The Standard
Model.

In the present paper we explore systematically the possibility, first
suggested in~\cite{Gri91}, that higgs bosons form bound states
(Higgsonium). In~\cite{Gri91}
we established, setting up a particular two-higgs model, that
non-relativistic loosely bound states of scalar bosons may indeed exist
for a wide range of masses and couplings in the scalar potential. And
this happens in a strictly perturbative regime, i.e. within the
conventional Standard Model framework.

Section~2 is devoted to the discussion of the mathematical techniques
used and the physical requirements needed for bound states to exist. In
section 3 we present the extended higgs sectors and the corresponding
lagrangians. An analysis of allowed regions in parameter space for bound
states to appear follows in section 4. Finally the last section  (5)
contains a summary of conclusions.

\section{Non relativistic bound states. The Yukawa Potential}

The Higgs potential in the standard model, or in any extension thereof,
contains cubic and quartic scalar couplings. In the standard model these
correspond to self-interactions of the higgs boson. In extended
scalar potentials they do also involve interactions among different
scalar bosons. In a two higgs model, for instance, with 5 physical
fields (3 neutral, 2 charged) triple quartic couplings among charged
and/or
neutral scalars are present in the lagrangian~(\cite{GHKD90}).
For certain domain of
physical parameters, the triple boson terms can be interpreted as a
classical density source of Yukawa field forces. As a result of
these Yukawa interactions, binding among (heavy) scalars may result
from exchange of a (light) scalar boson. This is the physical
situation which we shall investigate in this paper.

The general physical requirements we must impose are

\begin{enumerate}
\item[i)] The Yukawa potential produces a non-relativistic~(NR) bound state.

\item[ii)] The constituents of the bound states should outlive the bound-state
annihilation-time.

\item[iii)] The period of revolution of the bound system should be shorter
than both the lifetime of the constituents and the
annihilation time of the system.

\item[iv)] The repulsive quartic interactions which may prevent binding are
negligible.

\end{enumerate}

If those requirements are met, then higgsonium is possible.

A generic triple scalar coupling can be written,

\begin{equation}
        {\cal L}_3 =   - \frac{\rho_{ijk}}{n!}\; H^iH^jH^k ,
\end{equation}

with $H^{i,j,k}$ being identical or different fields (n is the number of
identical fields). The Yukawa potential corresponding to the particles
$i$ and $j$ exchanging particle $k$ is given by~\cite{Sak67},

\begin{equation}
        V(r) = - V_0 \frac{e^{-m_k r}}{m_k r},\label{eq:yuk1}
\end{equation}
with
\begin{equation}
        V_0 =  \frac{\rho_{iik}\rho_{jjk} M_k}{16 \pi M_i M_j}.
        \label{eq:yuk2}
\end{equation}

if particle $i$ is identical to particle $j$ then the potential is
necessarily attractive.

The potential generated by the 4-coupling
\begin{equation}
        {\cal L}_4 =  \frac{\rho_{ijkl}}{n!}\; H^iH^jH^kH^l
\end{equation}
is
\begin{equation}
V(\vec{r})=-\frac{\rho_{iijj}}{4M^2}\delta(\vec{r}).
\end{equation}
This is a contact potential. We shall estimate its influence on
higgsonium later.

If $V(r)$ is a Yukawa potential (or a sum of Yukawa potentials) then
the corresponding Schr\"odinger equation

\begin{equation}
  \left( -\frac{d^2}{2\mu dr^2} + \frac{l(l+1)}{2\mu r^2} + V(r) \right)
   u_{nl}(r) = E_{nl} u_{nl}(r),
\label{eq:sch1}
\end{equation}
with $\mu$ (the reduced mass) can be solved perturbatively by splitting
the potential in a Coulombic piece (The Yukawa potential closely
resembles  a Coulomb potential near the origin) plus the deviation from
an 1/r behaviour.

It is convenient to revert eq. (\ref{eq:sch1}) into an adimensional form by
dividing by a mass parameter squared:

\begin{equation}
  \left( -\frac{d^2}{dx^2} + \frac{l(l+1)}{x^2}
        -\frac{2g}{x} v(x) \right)
   \bar{u}_{nl}(x) = {\cal E}_{nl} \bar{u}_{nl}(x).
\label{eq:sch2}
\end{equation}

where the adimensional constant 2g characterizes the strength of the
potential. ${\cal E}_{nl}$ and $\bar u_{nl}(x)$ are the adimensional
eigenvalues and eigenfunctions.

This equation is then split in an hydrogen problem with "hamiltonian"
\begin{equation}
 H_0 = \left( - \frac{d^2}{dx^2} + \frac{l(l+1)}{x^2} - \frac{2g}{x} \right)
\end{equation}
plus a perturbation
\begin{equation}
       \Delta H = - \frac{2g}{x} ( v(x) - 1 ).
\end{equation}

First order perturbation theory gives,
\begin{equation}
  {\cal E}_{nl} = -\frac{g^2}{n^2} + \langle nl | \Delta H | nl \rangle
\end{equation}

where $\mid nl \rangle$ are hydrogen-like wave functions.

Next step is to expand $\Delta H$ in a Taylor series in $x$

\begin{equation}
  \Delta H = -2g \sum_{n=0}^{\infty} v_n x^n
\end{equation}
and realize that $\langle nl \mid \Delta H \mid nl\rangle$ can be written
in terms of the expectation values
\begin{equation}
  \langle nl|r^p|nl \rangle =  \langle p \rangle  g^{-p},
\end{equation}
where

\begin{equation}
  \langle p \rangle =
  \frac{n^{p-1}}{2^{p+1}}
  \frac{(n-l-1)!}{(n+l)!}
  \int_{0}^{\infty} dx e^{-x} x^{2+2l+p} \left(L_{n-l-1}^{2l+1}(x)\right)^2.
\end{equation}

finally, then
\begin{equation}
  {\cal E}_{nl}  =
  -g^2\left( n^{-2} + 2 v_0 g^{-1} +
	             2 v_1 \langle 1 \rangle g^{-2} +
                     2 v_2 \langle 2 \rangle g^{-3}
      \right) + O \left( 1/g^2 \right).
\label{eq:EnergiaPertorbativa}
\end{equation}

For a single Yukawa potential,
\begin{equation}
   v_n=\frac{(-1)^{n+1}}{(n+1)!} \hspace{1cm} {\rm and} \hspace{1cm}
   g=\frac{\mu V_0}{m^2}.
\end{equation}

This is an asymptotic series in $g^{-1}$ and gives accurate results for
large enough $g$, i.e. for a sufficiently strong potential. However,
for loosely bound systems like the ones  we shall encounter, the
convergence is improved notably~(\cite{GP75})
by the use of Pad\'e approximants~(\cite{Bak75}).
We include in appendix~\ref{ap:pade} a short account on the method.
It will permit us
to establish a threshold condition (a value for the parameters of the
potential) for a bound state to be formed (i.e. ${\cal E}_{nl}\leq 0)$.

The basic issue here is to find a threshold value for $g$, i.e. a value
above which the potential binds. From a numerical study close to this
threshold, both for $l=0$ and $l=1$, one concludes that optimal
Pad\'e approximants  for the energy are $\sqrt{-{\cal E}_{10}}^{[2/1]}$ and
$-{\cal E}_{nl}^{[2/1]}$ for S-wave and P-wave states respectively. The
explicit forms for these  Pad\'e approximants are given
in~(\ref{eq:Pade.2.1}), using the coefficients of
equation~(\ref{eq:EnergiaPertorbativa}). As an example, in the case of single
excange we obtain for the $l=0$ energy
\be
\sqrt{-\enertot{n0}}^{[2/1]}=\frac{g}{n}
\bfrac{ 1 + \frac{1-n^2}{3}\frac{1}{g} - \frac{n^2(4+5n^2)}{12} \frac{1}{g^2} }
{1+ \frac{1}{3} \frac{1+2n^2}{3} \frac{1}{g} },
\ee
and the threshold coupling $g$,
\be
g_{n0}^{[2/1]} = \frac{1}{6} \prt{ n^2-1 + \sqrt{16n^4 + 10n^2 +1} }.
\label{eq:Llindars}
\ee

The self-consistency of the approach requires that we deal with
non-relativistic systems, i.e. when binding is much smaller than
rest-mass:
\begin{equation}
                E_{nl} <\!\!< 2M
\end{equation}

This is physically achieved for small strength, large range~(light
exchanged mass) potentials.
As we shall explicitly  show, these conditions can be actually met in 2
higgs models.

We realized before that a contact potential might effect our bound
states. The perturbation induced on the $\vert nl\rangle$ states of higgsonium
by a potential of the form
\begin{equation}
     -\frac{\rho_{iiii}}{4M^2}\delta(\vec{r}).
\end{equation}
can be estimated to be

\begin{equation}
     \langle Contact \rangle \propto \rho_{iiii} \frac{m^3}{4M^2}
\end{equation}

for wavefunctions which correspond to potentials that are Coulombic
near the origin, i.e.

\begin{equation}
  \Psi_{nl} \propto m^{3/2} r^l \hspace{0.5cm}{\rm when}\hspace{0.5cm}
  r \rightarrow 0.
\end{equation}
This is to be compared with

\begin{equation}
  \langle Yukawa \rangle \propto \frac{m^2}{M}
  \label{eq:EstimacioYukawa}.
\end{equation}
The ratio
\begin{equation}
   \frac{\langle Contact \rangle}{\langle Yukawa \rangle}
   \propto \frac{m}{M}
\end{equation}
is small if we restrict ourselves to the nonrelativistic regime.

To comply with criteria {\em ii)} and {\em iii)} stated before it
is necessary now to introduce a ``characteristic time'' for the bound state.
We can give various definitions for this time, all differing only by a
numerical factor of order one. Perhaps the most obvious thing to do is to
define a ``classical revolution period'', i.e. the time it takes for the
constituents to complete a ``Bohr orbit'':

\begin{equation}
     \tau=2\pi \frac{\langle r \rangle}{\langle v \rangle}
     \label{eq:tau}
\end{equation}
with
\begin{equation}
  \langle r \rangle_{nl}= \langle nl | r | nl \rangle
  \hspace{0.5cm}{\rm and}\hspace{0.5cm}
  \langle v \rangle_{nl}= \frac{1}{m} \langle nl | p | nl \rangle
  \label{eq:ValorsEsperats}
\end{equation}

But, a perfectly sensible alternative would be to use the inverse of the
binding energy. The Heisenberg principle guarantees the
physical consistency of these characteristic times.

With such a time, one can then apply criteria {\em ii)} and {\em iii)} to
higgsonium. One merely requires that the width of the bound state and its
constituents is smaller than the inverse characteristic
time\footnote{whenever required we used Hulth\'en~\cite{HS57}
 wavefunctions to obtain
quantitative numerical estimates of equations such as~(\ref
{eq:EstimacioYukawa}), (\ref{eq:tau}) or (\ref{eq:ValorsEsperats}).}.

\section{Extended Higgs boson sectors}

The most simple and natural extension of the Standard Model involves
two complex higgs scalar $SU(2)_L$ doublets $\Phi_1$ and $\Phi_2$ with
hyperchange $Y=1/2$.

The most general higgs potential -gauge invariant, CP-conserving and
renormalizable- with automatic exclusion of FCNC is,

\begin{eqnarray*}
\lefteqn{V(\phi_1,\phi_2) =} \\
  & &
\mu_1^2     {\phi}_1^{\dagger}\phi_1 +
\mu_2^2     \phi_2^{\dagger}\phi_2 +
               \lambda_1 ( \phi_1^{\dagger}\phi_1 )^2 +
               \lambda_2 ( \phi_2^{\dagger}\phi_2 )^2 +\\
  & &
               \lambda_3  ( \phi_1^{\dagger}\phi_1)(\phi_2^{\dagger}\phi_2) +
               \lambda_4  ( \phi_1^{\dagger}\phi_2)(\phi_2^{\dagger}\phi_1) +
               \lambda_5  [( \phi_1^{\dagger}\phi_2)^2 +
                           ( \phi_2^{\dagger}\phi_1)^2 ]
\end{eqnarray*}
where all constants are real.

In addition, this potential is bounded below if
\begin{equation}
\begin{array}{clclclc}
\lambda_1>0  & , &
\lambda_2>0  & , &
\lambda_3 + \lambda_4 + \lambda_5 > - 2 \sqrt{ \lambda_1 \lambda_2} & , &
\lambda_4 + \lambda_5<0 \\
\end{array}
\end{equation}

Spontaneous breaking of the electroweak symmetry down to
electromagnetism occurs when $\phi_1$ and $\phi_2$ acquire v.e.v's $v_1$
and $v_2$ (which we take both real, since we are not interested in
spontaneous CP violation). In terms of these vev's we rewrite the
fields as,

\begin{equation}
\begin{array}{cc}
        \phi_1 = \frac{1}{\sqrt{2}}     \left(
                                        \begin{array}{c}
                                                \phi_1^{+}\\
                                                v_1+R_1+i I_1\\
                                        \end{array}
                                        \right)
&
        \phi_2 = \frac{1}{\sqrt{2}}     \left(
                                        \begin{array}{c}
                                                \phi_2^{+}\\
                                                v_2+R_2+i I_2\\
                                        \end{array}
                                \right)
\end{array}
\end{equation}
where $R_i={\rm Re} \phi_i^0-v_i$ and $I_i={\rm Im} \phi_i^0$.

Also,

\begin{eqnarray}
\mu_1^2 &=& -\lambda_1 v_1^2 -
            \frac{1}{2} (\lambda_3+\lambda_4+\lambda_5) v_2^2, \nonumber \\
\mu_2^2 &=& -\lambda_2 v_2^2 -
            \frac{1}{2} (\lambda_3+\lambda_4+\lambda_5) v_1^2.
\end{eqnarray}

The physical degrees of freedom, in terms of the 8 real fields in
$\phi_1$ and $\phi_2$ are explicitly given by,

\begin{eqnarray}
       H^\pm & = & - \phi_1^\pm \sin \beta + \phi_2^\pm \cos \beta
\end{eqnarray}
i.e. two charged higgs bosons, and
\begin{eqnarray}
  A^0 & = &  -  I_1 \sin \beta + I_2 \cos \beta
\end{eqnarray}
a pseudoscalar neutral boson and, finally, the neutral scalars
\begin{eqnarray}
   H^0 & = & \mbox{ } R_1 \cos \alpha + R_2 \sin \alpha \\
   h^0 & = &    -     R_1 \sin \alpha + R_2 \cos \alpha,
\end{eqnarray}
with

\begin{equation}
\begin{array}{clc}
 \tan \beta  = v_2 / v_1 &,& tan 2 \alpha = \frac{C}{A-B}
\end{array}
\end{equation}
\begin{equation}
\begin{array}{clclc}
       A=v_1^2 \lambda_1   &, &
       B=v_2^2 \lambda_2   &, &
       C=v_1 v_2 (\lambda_3+\lambda_4+\lambda_5)\\
\end{array}
\end{equation}

Their masses are:
\begin{eqnarray}
m_{H^\pm}^2   & = & -\frac{1}{2} (v_1^2+v_2^2)(\lambda_4+\lambda_5)\nonumber \\
m_{A^0}^2     & = & -\lambda_5 (v_1^2+v_2^2) \nonumber \\
m_{H^0,h^0}^2 & = & A+B \pm \sqrt{(A-B)^2+C^2}
\end{eqnarray}

For convenience of analysis, in section 4, we shall trade the 7
parameters $\mu_1, \mu_2, \lambda_{1,2,3,4,5}$ for the 7 related
ones: the masses $m_{H^\pm}, m_{A^0}, m_{H^0}, m_{h^0}$, the mixing
angles $\alpha$ and $\beta$, and the vev
$v=\sqrt{\frac{1}{2}(v_1^2+v_2^2)}$.

Finally, to specify completely  our  model, we take
doublet $\phi_1$ to be coupled to all down-type right-handed~(RH) fermions
and hence
gives masses to $d, s, b$ quarks and to the charged leptons ($e$, $\mu$,
$\tau$). Similarly, the higgs field $\phi_2$ is coupled to up-type
($u$,$c$,$t$) RH fermion fields  and is responsible for their masses.

The neutral higgs bosons will have to be more massive than 60 GeV. The
charged higgs boson masses should not lie below 45 GeV.

As to tan $\beta$, we will require
\begin{eqnarray}
0.7 &<\tan\beta<& 50
\end{eqnarray}

This comes about from the requirement that the higgs boson couplings to
the fermions are perturbative (of course, the actual restrictions arise
from the top and bottom quarks).

\section{Results}

The states that can be a priori constructed  (the forces being
necessarily attractive) are
$\langle H^+ H^- \rangle ,
 \langle A^0 A^0 \rangle ,
 \langle h^0 h^0 \rangle $ and
$\langle H^0 H^0 \rangle$. However,
$\langle h^0 h^0 \rangle$ and
$\langle H^0 H^0 \rangle$ would  not qualify as
non-relativistic systems since $m_{H^0}\!\!>\!m_{h^0}$ and $H^0$ is the
particle to be exchanged. We therefore shall study the S-wave states
$\langle H^+ H^- \rangle_{10}$
and
$\langle A^0 A^0 \rangle_{10}$.

We have analyzed also the P-wave states, but due to the centrifugal
barrier their formation is not possible.

\subsection{The state $\langle H^+ H^-\rangle_{10}$}

This state can be formed by the exchange of $H^0$ and $h^0$ (the vertex
$H^+H^-A^0$ does not exist) whose Feynman rules are collected in
appendix B.

To cope with a 6-parameter analysis we shall in what follows choose
definitive sets of masses and plot in the $(\alpha, tan \beta)$-plane
the corresponding allowed regions where the state
$\langle H^+H^-\rangle$ is
possible, i.e. where ${\cal E}_{10}\!\!<\!0$ and criteria {\em i)},
{\em ii)} and {\em iii)} are fullfilled.

The choice of masses is dictated by the requirements of producing
non-relativistic systems (constituent mass much larger than
exchanged mass), exceeding the lower bounds
of $45GeV$ and $60GeV$ for charged and neutral scalars respectively, and
complying with unitarity (i.e. scalar masses below about 1000 GeV).

In order to compare the ``orbiting time'' with the lifetime of the
constituents $H^{\pm}$ we need to compute the decays of the charged
higgs boson. Appendix C  is a full listing of the partial widths of
$(H^\pm, H^0, A^0, h^0)$. From these formulae we easily calculate the
relevant $H^{\pm}$ total width.

The lifetime of the $\langle H^+H^- \rangle_{10}$ state is
obtained~\cite{Jac76} from the width,

\begin{equation}
\Gamma\left(\langle H^+ H^-\rangle_{10}\right) =
                                |\Psi(0)|^2 v_{rel} \;
                                \sigma(H^+ H^- \rightarrow all),
\end{equation}
where
\begin{equation}
\sigma(H^+ H^- \rightarrow all) =
\sigma(H^+H^- \rightarrow H^iH^j, t\bar{t}, W^+W^-, ZZ, \cdots)
\end{equation}

For the wave function at the origin we use the Hulth\'en wave function.

In figures
(\ref{fig:pm300}), (\ref{fig:pm400}) and (\ref{fig:pm500})
we display our numerical results, for the
constituent masses of 300 GeV, 400 GeV and 500 GeV, respectively, and
the neutral higgs boson masses shown. We see clearly
that there is much parameter space to allow for higgsonium.

\subsection{The state $\langle A^0A^0 \rangle_{10}$}

It can be formed via the exchange of $H^0$ and $h^0$ bosons. The
corresponding couplings are explicitly given by eqs. B.7 and B.8.

In our analysis we shall demand that $m_{A^0}$ is much larger than
$m_{H^0}$ and $m_{h^0}$ (recall the non-relativistic character
of the approach).

The construction of the allowed regions in parameter space follows the
same strategy as for the $\langle H^+H^- \rangle_{10}$-states.

The results are shown in
figures
(\ref{fig:aa400}) and (\ref{fig:aa500})
( for $m_{A^0}=$ 400 GeV and 500 GeV,
respectively).

Again, formation of higgsonium $\langle A^0A^0 \rangle_{10}$ is possible,
at least for a considerable range of higgs boson masses and mixing angles.

\subsection{The MSSM case}
In the MSSM~(\cite{HK85}),
the extra constraints imposed by supersymmetry on the scalar
potential lead to the following couplings:

\ba
\rho^{\scriptscriptstyle M\!S\!S\!M}_{A^0A^0H^0} & = &
        \bfrac{ig\MZ}{2\cw}\cos2\beta \; \cos(\beta+\alpha) \\
\rho^{\scriptscriptstyle M\!S\!S\!M}_{A^0A^0h^0} & = &
        \bfrac{ig\MZ}{2\cw}\cos2\beta \; \sin(\beta+\alpha)\\
\rho^{\scriptscriptstyle M\!S\!S\!M}_{H^+H^-H^0} & = &
        -ig\prt{ \MW \cos(\alpha-\beta) -
	         \frac{\MZ}{2\cw} \cos2\beta\cos(\alpha+\beta)}\\
\rho^{\scriptscriptstyle M\!S\!S\!M}_{H^+H^-H^0} & = &
        -ig\prt{ \MW \sin(\alpha-\beta) +
                 \frac{\MZ}{2\cw} \cos2\beta\sin(\alpha+\beta)
               }.
\ea

These couplings should be responsible for the formation of the
$\langle H^+H^- \rangle_{10}$ and $\langle A^0A^0 \rangle_{10}$
bound states. However, using the techniques explained before, we reach
the conclusion thath their strength is not sufficient for these bound
states to exist. Therefore, the MSSM does not produce higgsonium.

\section{Summary}

The Higgs boson is the only element of the Standard Model yet to be
disclosed by experiment. In the minimal version of the standard
electroweak model, only one neutral Higgs scalar exists, but non
minimal extensions are possible and even theoretically desirable. Indeed
more ambitious constructs such as Grand Unified Theories or
Supersymmetry demand larger scalar sectors.
The most natural extension of the minimal scalar sector is to include
two $SU(2)_L$ doublets. This extension implies the existence of two
charged scalars particles and three neutral particles.

In this paper we have considered a generic two higgs doublet model and
explored the possibilities that, for certain choices of the parameters
in the scalar potential, bound states of Higgs scalars are produced.
This should come about through the appearence of attractive Yukawa-type
forces among identical Higgs particles, charged or neutral. We have
shown that bound states of the systems $H^+H^-$ and $A^0A^0$ in an
S-wave mode are indeed possible for a wide range of parameters in the
Higgs potential. Also, P-wave or higher angular momentum states are
ruled out in 2-Higgs models. Since the Minimal Supersymmetric Standard
Model is, as far as its Higgs sector is concerned, a model in the
generic class of two Higgs models one might wonder whether scalar bound
states can be achieved in this case. It turns out that this is not the
case and the reason lies in the strong extra constraints imposed by
Supersymmetry. In this case, the 6 parameter space ($m_{H^\pm}$,
$m_{H^0}$, $m_{h^0}$, $m_{A^0}$, $\alpha$, and $\tan\!\beta$) reduces
to a 2 parameter space: $\tan\!\beta$ and one scalar particle mass
(e.g. $m_{A^0}$).

\vspace{3cm}
This work has been partially supported by CICYT under project under
AEN-93-0474

\newpage

\appendix
\section{The Pad\'e approximants}
\label{ap:pade}


Let $A(x)$ be a  function whose formal series expansion is

\begin{equation}
	A(x) = \sum_{j=0}^{\infty} a_j x^j.
\end{equation}

We define

\begin{equation}
	A^{[L/M]}(x) = \bfrac{P_L(x)}{Q_M(x)}
\end{equation}

as the $[L/M]$ Pad\'e approximant of function $A(x)$, where $P_L(x)$
is a polynomial of order less or equal than $L$ and $Q_M(x)$ is a
polynomial of order less or equal than $M$. $P_L(x)$ and $Q_M(x)$
verify:

\begin{enumerate}

\item	$\sum_{j=0}^{\infty} a_j x^j - A^{[L/M]}(x) = O(x^{L+M+1})$.

\item	$Q_M(0) = 1$.

\item	$P_L$ and $Q_M$ do not have common factors.

\end{enumerate}

\begin{theo}
The Pad\'e approximant $[L/M]$ to a formal series expansion (if it
exists) is unique.
\end{theo}

This theorem allows the explicit construction of the various
approximants:

\begin{eqnarray}
\label{eq:Pade.0.0}
[0/0] &=& a_0						\\[0.2cm]
\label{eq:Pade.1.0}
[1/0] &=& a_0 + a_1 x					\\[0.2cm]
\label{eq:Pade.0.1}
[0/1] &=& \frac{a_0}{(a_1/a_0)x}			\\[0.2cm]
\label{eq:Pade.1.1}
[1/1] &=& \frac{a_0+\left(a_1-a_0 \; a_2/a_1\right) x}
		{1-(a_2/a_1)x}				\\[0.2cm]
\label{eq:Pade.2.0}
[2/0] &=& a_0 + a_1 x + a_2 x^2				\\[0.2cm]
\label{eq:Pade.2.1}
[2/1] &=& \frac{a_0 + \left( a_1-a_0 \; a_3/a_2 \right) x +
		      \left( a_2-a_1 \; a_3/a_2 \right) x^2
		}
		{1-(a_3/a_2)x}			\\[0.2cm] \nonumber
& \vdots &
\end{eqnarray}

\begin{conj}[Pad\'e]
If a function $F(z)$ is regular inside a circle $|z|\!\!\!<\!\!\!R$
except for
$m$ poles within this circle, then there exist at least a subsequence
of the diagonal Pad\'e approximants which converges uniformly to
$F(z)$ inside the domain obtained by removing from this circle the
interior of small circles centered at the poles.
\end{conj}

This conjecture allows to obtain large convergence regions for large
classes of functions. In practice one can test it on numerical
examples.

\section{Feynman rules involving the Higgs bosons}

\subsection{Gauge bosons and Higgs bosons}

$$
\begin{array}{ll}
  \rfa  \scriptstyle{ie(p-q)_{\mu}}
& \rfb  \scriptstyle{i\frac{g}{2\sw\cw}(\cws-\sws)(p-q)_{\mu}}
\\[1ex]
  \rfc  \scriptstyle{\frac{g\cos(\ab)}{2\cw}(p-q)_{\mu}}
& \rfd  \scriptstyle{-\frac{g\sin(\ab)}{2\cw}(p-q)_{\mu}}
\\[1ex]
  \rfe  \scriptstyle{\frac{g}{2}(p-q)_{\mu}}
& \rff  \scriptstyle{-i\frac{g}{2}\cos(\ab)(p-q)_{\mu}}
\\[1ex]
  \rfg  \scriptstyle{i\frac{g}{2}\sin(\ab)(p-q)_{\mu}}
& \rfh  \scriptstyle{i\frac{g\MZ}{\cw}[\cos(\ab),\sin(\ab)]g_{\mu\nu}}
\end{array}
$$

$$
\begin{array}{ll}
  \rfi  \scriptstyle{igM_W[\cos(\ab),\sin(\ab)]g_{\mu\nu}}
& \rfj  \scriptstyle{i\frac{ge}{\cw}(\cws-\sws)g_{\mu\nu}}
\\[1ex]
  \rfk  \scriptstyle{i\frac{g}{4\cws}(\cws-\sws)g_{\mu\nu}\,2!}
& \rfl  \scriptstyle{i\frac{g^2}{2}g_{\mu\nu}}
\\[1ex]
  \rfm  \scriptstyle{i\frac{g^2}{8\cws}g_{\mu\nu}\,2!\,2!}
& \rfn  \scriptstyle{i\frac{g^2}{4}g_{\mu\nu}\,2!}
\end{array}
$$
\subsection{Fermions and Higgs bosons}

$$
\begin{array}{rllr}
\sfrac{-igM_t\sin\alpha}{2\MW\cos\beta} & \figutt &
\sfrac{-igM_b\cos\alpha}{2\MW\cos\beta} & \figubb
\\[1ex]
\sfrac{-igM_t\cos\alpha}{2\MW\sin\beta} & \figltt &
\sfrac{-igM_b\sin\alpha}{2\MW\cos\beta} & \figlbb
\\[1ex]
\sfrac{-gM_t\cot\beta}{2\MW}  & \figatt &
\sfrac{-gM_b\cot\alpha}{2\MW} & \figabb
\\[1ex]
\end{array}
$$
$$
\begin{array}{rlrl}
\begin{array}{l}
        \sfrac{ig}{2\sqrt{2}\MW} \\
        \scriptstyle{
        \left[(M_b \tan\beta +M_t\cot\beta)+\right.}\\
        \scriptstyle{
        \left. (M_b \tan\beta -M_t\cot\beta)\gamma_5 \right]}
\end{array} &
\figptb &
\begin{array}{l}
        \sfrac{ig}{2\sqrt{2}\MW}\\
        \scriptstyle{
        \left[(M_b \tan\beta +M_t\cot\beta)-\right.}\\
        \scriptstyle{
        \left. (M_b \tan\beta -M_t\cot\beta)\gamma_5 \right]}
\end{array} &
\figmbt
\end{array}
$$

\subsection{3-scalar vertices}

\begin{eqnarray}
  \rho_{ H^+ H^- H^0 } & = & \eqhpmu \\
  \rho_{ H^+ H^- h^0 } & = & \eqhpml \\
  \rho_{ H^0 H^0 H^0 } & = & \eqhuuu \\
  \rho_{ h^0 h^0 h^0 } & = & \eqhlll \\
  \rho_{ H^0 H^0 h^0 } & = & \eqhuul \\
  \rho_{ h^0 H^0 H^0 } & = & \eqhllu \\
  \rho_{ A^0 A^0 H^0 } & = & \eqhaau \\
  \rho_{ A^0 A^0 h^0 } & = & \eqhaal
\end{eqnarray}

\subsection{4-scalar vertices}

\eqquatrehiggs

Notice that vertices with an odd number of $A^0$'s vanish (CP conservation)

\section{Partial widths of Higgs bosons}

\subsection{Fermionic decays}
\label{sec:ToFermions}

To be definite, we display our formulae in terms of the 3rd family of quarks.

\ba
  \Gamma(H^+ \rightarrow t\bar{b}) & = &
	  \bfrac{3 g^2 \lambda^{1/2}}{32 \pi \MW^2 M_{H^+}^3}
	   \left( (M_{H^+}^2-M_b^2-M_t^2) \right. \nonumber\\
	  & &
	  \left. (M_b^2\tan^2\beta +M_t^2\cot^2\beta) -4M_b^2 M_t^2 \right)
\label{eq:pmtotb} \\[1cm]
  \Gamma(H^i \rightarrow t\bar{t}) & = &
	\bfrac{3 g^2 M_t^2 u_i^2 M_i}{32 \pi \MW^2 \sin^2\beta}
	  \prt{1-\frac{4M_t^2}{M_i^2}}^p
\label{eq:iitott}\\[1cm]
  \Gamma(H^i \rightarrow b\bar{b}) & = &
        \bfrac{3 g^2 M_b^2 d_i^2 M_i}{32 \pi \MW^2 \cos^2\beta}
          \prt{1-\frac{4M_t^2}{M_i^2}}^p
\label{eq:iitobb}
\ea
where

\be
u_i  \equiv \left\{  \begin{array}{rl}
              -\sin\alpha,	& H^i=H^0 \\
              \cos\alpha,	& H^i=h^0 \\
              \cos\beta,	& H^i=A^0
              \end{array}
      \right. ,\\
\ee
and
\be
d_i  \equiv \left\{  \begin{array}{rl}
              \cos\alpha,	& H^i=H^0 \\
              \sin\alpha,	& H^i=h^0 \\
              -\sin\beta,	& H^i=A^0
              \end{array}
      \right. .
\ee
Also
\be
   p \equiv \left\{  \begin{array}{ll}
              3/2,       & H^i=H^0,h^0 \\
              1/2,       & H^i=A^0
              \end{array}
      \right. .
\ee
and
\be
	\lambda^{1/2} \equiv
		\cld{(M_1^2 + M_2^2 - M_3^2)^2 - 4 M_1^2 M_2^2 }^{1/2},
\ee
($M_1$ $M_2$ i $M_3$ are the masses of the particles participating in the
decay).

For decays into leptons, one should suppress the color factor 3.

\subsection{Decays into a pair of gauge bosons}

\label{sec:ToGaugeBosons}

\ba
  \Gamma(H^i \rightarrow W^+W^-) & = &
	\bfrac{g^2 (M_{H^i}^4-4M_{H^i}^2\MW^2+12\MW^4) V^i}
              {64 \pi \MW^2 M_{H^i}}
	  \prt{1-\frac{4\MW^2}{M_{H^i}^2} }^{1/2}
	  \label{eq:HitoWW}\\[1cm]
  \Gamma(H^i \rightarrow ZZ) & = &
	\bfrac{ g^2 (M_{H^i}^4-4M_{H^i}^2\MZ^2+12\MZ^4) V^i}
              {128 \pi \MZ^2 \cws M_{H^i} }
	  \prt{1-\frac{4\MW^2}{M_{H^i}^2} }^{1/2}
          \label{eq:HitoZZ}
\ea
with

\be
V^i  \equiv \left\{  \begin{array}{cl}
              \cos^2\alpha,	& H^i=H^0 \\
              \sin^2\alpha,	& H^i=h^0 \\
              0,		& H^i=A^0
              \end{array}
      \right. .
\ee

\subsection{Decays into a gauge boson and a Higgs boson}

\be
  \Gamma(H^\pm \rightarrow W^\pm H^i) =
       \bfrac{ g^2 \lambda^{1/2} G^i}{64 \pi \MH^3}
       \cld{ \MW^2 -2(\MH^2+M_{H^i}^2)+\bfrac{(\MH^2-M_{H^i}^2)^2}{\MW^2}},
\label{eq:HtoWHi}
\ee
where

\be
G^i  \equiv \left\{  \begin{array}{cl}
              \sin^2\alpha,	& H^i=H^0 \\
              \cos^2\alpha,	& H^i=h^0 \\
              1,		& H^i=A^0
              \end{array}
      \right. .\\
\ee

\be
  \Gamma(H^i \rightarrow Z H^j)=
       \bfrac{ g^2  \lambda^{1/2} G^{ij}}{64 \pi M_{H^i}^3}
       \cld{ \MZ^2 -2(M_{H^i}^2+M_{H^j}^2)+
				\bfrac{(M_{H^i}^2-M_{H^j}^2)^2}{\MZ^2}},
\label{eq:HitoZHj}
\ee
where
\be
G^{ij}  \equiv \left\{  \begin{array}{cl}
              \sin^2\alpha,     & H^i,H^j=H^0,A^0 \\
              \cos^2\alpha,     & H^i,H^j=h^0,A^0 \\
              0,                & H^i,H^j=H^0,h^0
              \end{array}
      \right. .\\
\ee

\subsection{Decays into two Higgs scalars}

Whenever kinematically possible:
\ba
\Gamma(H^i \rightarrow H^j H^k)&=& \bfrac{1}{4\pi M_{H^i}}
      \prt{ 1- (M_{H^j}+M_{H^k})^2/(4M_{H^i}^2) }^{1/2}\nonumber\\
  & & \prt{ 1- (M_{H^j}-M_{H^k})^2/(4M_{H^i}^2) }^{1/2}
  \rho_{ijk} \frac{1}{(1+\delta_{jk})!},
\label{eq:HitoHjHk}
\ea
where $\rho_{ijk}$ is given by equations (\ref{eq:hpmu})-(\ref{eq:haal}).

\newpage
\figpmtrescents
\figpmquatrecents
\figpmcinccents
\figaaquatrecents
\figaacinccents

\end{document}